\documentclass[manuscript,screen,nonacm]{acmart}
\setcopyright{rightsretained} 
\copyrightyear{2025}
\acmJournal{POMACS}

\received{February 2025}

\usepackage[none]{hyphenat}
\usepackage{balance}
\usepackage[utf8]{inputenc}
\usepackage[T1]{fontenc}
\usepackage{dirtytalk}
\usepackage{xcolor,colortbl}
\usepackage{graphicx}
\usepackage{libertine}
\usepackage{microtype}
\usepackage{ccicons}
\usepackage{ulem}[normalem]
\usepackage{textcomp}
\usepackage{etoolbox}
\usepackage{booktabs}
\useunder{\uline}{\ul}{}
\usepackage{url}

\definecolor{Highlight}{rgb}{0.0,0.0,0.0}
\definecolor{Highlight2}{rgb}{0.0,0.0,0.0}
\definecolor{rv1}{rgb}{0.0,0,0.0}
\definecolor{g6}{gray}{1.00}
\definecolor{g5}{gray}{0.94}
\definecolor{g4}{gray}{0.88}
\definecolor{g3}{gray}{0.81}
\definecolor{g2}{gray}{0.73}
\definecolor{g1}{gray}{0.65}
\definecolor{g0}{gray}{0.54}
\definecolor{sd}{gray}{0.54}
\definecolor{NA}{rgb}{0.6,0.4,0.4}
\usepackage{fancyhdr}
\usepackage{cuted}
\usepackage{capt-of}
\usepackage{hyperref}
\definecolor{mygray}{gray}{0.6}
\makeatletter
\def\url@leostyle{  \@ifundefined{selectfont}{
    \def\UrlFont{\sf}
  }{
    \def\UrlFont{\small\bf\ttfamily}
  }}
\makeatother
\graphicspath{{figs/}}

\def\plaintitle{Game Changers: Designing and Measuring Dynamic Feedback To Help Users Self-Regulate in a VR Pointing Game}

\usepackage[resetlabels]{multibib}
\newcites{game}{Ludography}
\newcommand{\citegameprefix}{G}
\usepackage{xparse}

\let\origcitegame\citegame
\RenewDocumentCommand{\citegame}{O{} O{} m}{  \renewcommand{\citenumfont}[1]{\citegameprefix##1}  \origcitegame[#1][#2]{#3}  \renewcommand{\citenumfont}[1]{##1}}

\def\plainauthor{Bastian Ilsø Hougaard et al.} 

\def\plainkeywords{Serious Games; Feedback Timing; Intelligent Feedback; Core Tasks; Game Assistance;}

\begin{document}

\title[Designing and Measuring Dynamic Feedback To Help Users Self-Regulate]{\plaintitle}

\author{Bastian~Ils\o~Hougaard}
\orcid{0000-0002-6861-1858}
\email{biho@create.aau.dk}
\affiliation{  \institution{Aalborg~University}
  \city{Aalborg}
  \country{Denmark}}
\author{Scott Bateman}
\orcid{0000-0003-3592-2163}
\affiliation{  \institution{University of New Brunswick}
  \city{Fredericton}
  \country{New Brunswick, CA}}
\author{Iris Brunner}
\orcid{0000-0002-7194-3087}
\affiliation{  \institution{Aarhus~University}
  \city{Aarhus}
  \country{Denmark}}
\author{Lars Evald}
\orcid{0000-0001-6255-2562}
\affiliation{  \institution{Aarhus~University}
  \city{Aarhus}
  \country{Denmark}}
\author{Hendrik~Knoche}
\orcid{0000-0003-3950-8453}
\affiliation{  \institution{Aalborg~University}
  \city{Aalborg}
  \country{Denmark}}

\renewcommand{\shortauthors}{Hougaard et al.}

\begin{abstract}
The way games dynamically convey information through feedback is critical to players' ability to perform, learn, and improve.
However, it is poorly understood how performance metrics impact player performance and perception in core game tasks like pointing or steering. 
{\color{rv1} With a virtual reality pointing task we systematically explored how three performance metrics driving the feedback affected players when 
rewarding short completion times, straight movements, or high peak speed. across different points in time - continuously, at end-of-action, or at end-of-task.}
On average the dynamic feedback helped people point more straight and faster, while for others it had small or opposite effect. 
{\color{rv1}The study quantitatively compared dynamic feedback across three forms with the metrics driving the form as the intended locus of quantitative comparison.}
Our work improves game designers basis for crafting dynamic feedback by helping them know when to employ feedback schemes that align with desirable game performance objectives.
\end{abstract}


\definecolor{linkColor}{RGB}{6,125,233}
\hypersetup{
  pdftitle={\plaintitle},
  pdfauthor={\plainauthor},
  pdfkeywords={\plainkeywords},
  bookmarksnumbered,
  citecolor=black,
  filecolor=black,
  linkcolor=black,
  urlcolor=linkColor,
  breaklinks=true,
hypertexnames=false}


\begin{teaserfigure}
      \centering
      \includegraphics[width=0.90\textwidth]{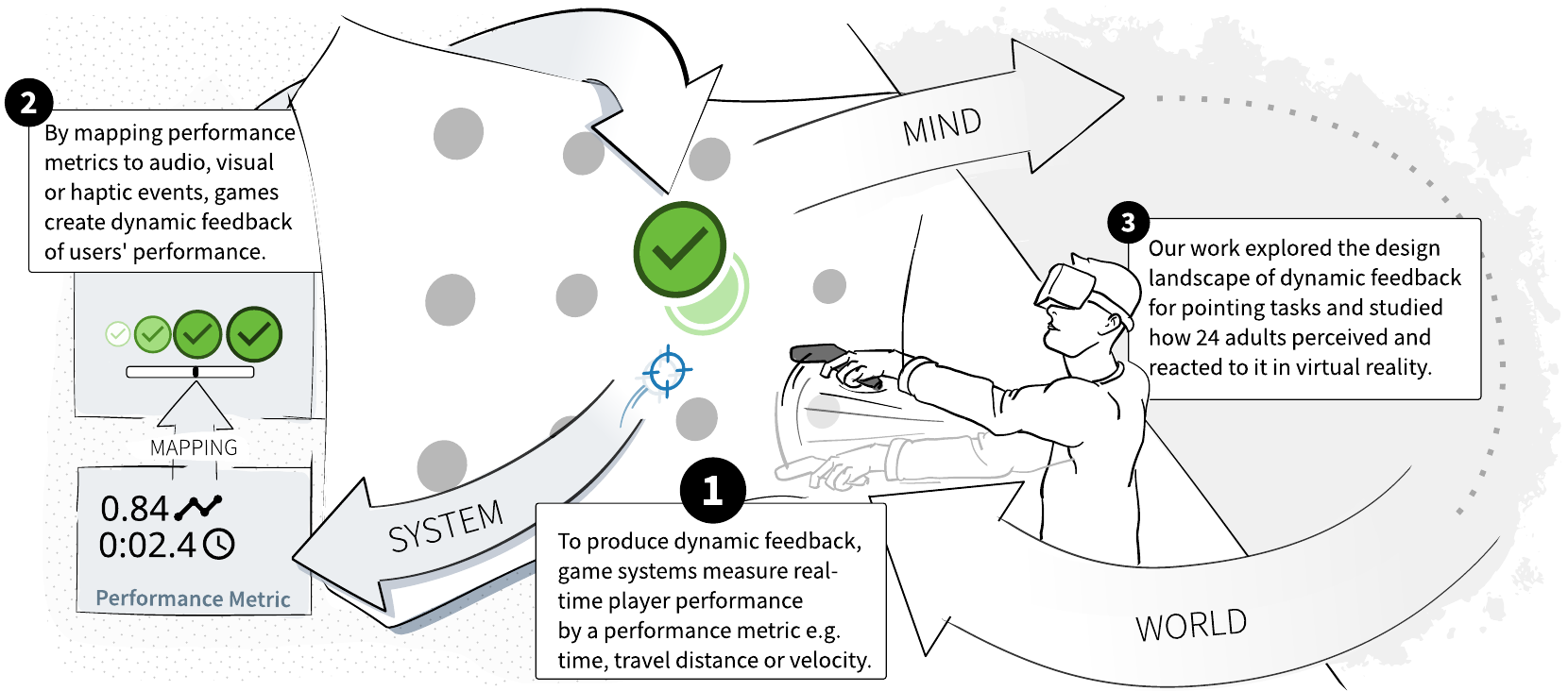}
      \caption{Our study explores the continuous self-regulatory loop in virtual reality between a player's movement performance changing a pointing game's haptic, visual and auditory dynamic feedback (e.g. checkmarks with dynamically varying intensity).
            }
      \label{fig:conceptualspace}
      \Description{A drawn person, seen from the side, is wearing a head-mounted display and moves a controller around to point a blue crosshair at a green target with their left arm. The person is surrounded by a virtual world containing a large wall of targets, annotated by a loop of arrows and boxes with text. An arrow titled system, points from the moving crosshair to a textbox tiled metric, described as step 1: to produce dynamic feedback, game systems measure real-time player performance by a performance metric e.g. time, travel distance or velocity. The metric textbox points to a textbox behind the wall titled feedback, described as step 2: By mapping performance metrics to audio, visual  or haptic events, games create dynamic feedback of users' performance.  A large arrow points from a checkmark on the wall into the person's mind described as step 3: In this work, we explore the design landscape of dynamic feedback, and study how users'perceive and react to it. From here a new arrow pointing back into the virtual world at the person's moving arm.}
\end{teaserfigure}

\maketitle

\section{Introduction}
In human-in-the-loop systems, \textit{feedback} describes the sensory elements designed to inform users about their interactions with the system. 
For example, a clicking sound on button press encodes the information that the system has registered the user’s input.
Games are commonly designed with an abundance of responsive elements that encode important information, and they are used to not only support an understanding of the game's state but also maximize players' perception of rewards and steer performance. Within game interaction loops, feedback typically consists of the visual, auditory, and haptic information displayed to players, influencing their understanding and enjoyment of the game experience~\citep{Schell_2019}.
The merit of any information-display mapping can be considered from players' ability to utilize them to improve their performance and in terms of \textit{correspondence} – their perception of how well the game's feedback reflects their performance.
The quality of an information-display mapping depends on how well players can use it to improve their performance and their perception of how accurately the game's feedback reflects their actions.
For example, in \textit{Fruit Ninja}~\citep{halfbrick2010fruitninja}, the speed of players’ strokes corresponds to larger and louder swings of their swords, creating a rewarding experience. However, it is unclear whether such feedback is merely an audio-visual embellishment that varies feedback to reduce monotony 
or whether  communicating measurements in this indirect way can affect players' performance positively or negatively. 
With the right feedback, game designers could guide Fruit Ninja players to make quicker, more expansive movements, enhancing their chances of successfully slicing multiple fruits.
However, players might instead prioritize speed over accuracy, either in pursuit of satisfying feedback or because they interpret the feedback as endorsing their actions and thereby impair their own  performance.

This highlights a knowledge gap for game designers, as the vast range of possible information-display mappings remains poorly understood in terms of their effects on player performance and behavior.  
To date, questions relating to information-display mappings have been addressed by relying on play-testing, gut feeling, or copying known practices. Game scholarship has contributed with design lenses to aid feedback design, like Hick’s ``juicy'' design framework~\citep{HicksFramework2018}, Deterding’s feedback design lenses~\citep{Deterding_2015}, and Schell’s lens of feedback, juiciness, channels, and information~\citep{Schell_2019}. 
However, although juicy feedback can demonstrably benefit task performance in non-game scenarios~\citep{sigrist_augmented_2013}, game studies of juicy feedback suggested that it may not improve performance and may even worsen outcomes compared to minimal non-juicy feedback~\citep{Hicks_2019,Juul2016GoodFF}. 
Game designers and researchers are missing fundamental knowledge of the extent players can use information-display mappings to improve performance. Including the answers to questions like: How much do users read into the feedback received and use it to improve their performance? When will feedback induce a cognitive cost that affects performance negatively? 

To answer such questions, 
we designed a study that evaluated players' performance and experience with three performance metrics implemented into feedback happening at different points in time (from immediate feedback to more periodic feedback). Our study in Section~\ref{sec:feedback-study} consists of the following contributions:
\begin{enumerate}
    \item We unpack the design space of dynamic feedback by combining Wensveen's couplings of time and dynamic with concepts from motor learning and game scholarship  (Section~\ref{sec:dimensionsfeedback}).
    \item We conceptualize a self-regulation model for pointing games, and draw upon information processing theory to propose how to measure key experiential dimensions affected by feedback design choices (Section~\ref{sec:conceptual-model}).
    \item We designed a low-level action-feedback game loop to study users' ability to improve performance while receiving dynamic feedback based on three distinct metrics (Section~\ref{sec:feedback-study}). 
    \item  We provide empirical evidence linking users' performance self-regulation to their overall performance. Our qualitative results show factors affecting dynamic feedback's legibility (Section~\ref{sec:results}). 
\end{enumerate}

Our systematic feedback design approach for pointing tasks (Section~\ref{sec:dimensionsfeedback}) can optimize feedback in a wide range of applied pointing-based games, including motion-controlled games, exergames, and rehabilitation games. 
Based on our quantitative analysis of tracked participant movements, self-report measurements, and semi-structured interviews with participants, we provide design recommendations for pointing tasks employing audiovisual real-time feedback.
With the growing interest in designing video games for complex scenarios from rehabilitation and sport training to the use of artificial intelligence (AI), 
it becomes increasingly important to understand how feedback guides players' game behaviour. 
Our study reveals valuable insights into how low and high-performing players derive benefits from dynamic, performance metric-driven feedback.

\section{Background}
In HCI, feedback constitutes a fundamental concept in interaction and product design, complementing notions like feedforward, signifiers, and affordance.
Within game scholarship feedback closely relates to design lenses like \textit{game feel}~\citep{Swink_2014} and game-specific terminology like \textit{juiciness}~\citep{Schell_2019,Deterding_2015} -- a reference to non-functional feedback designed to evoke a visceral feeling through an abundance of (typically) visual effects~\citep{Hicks_2019,Schell_2019}. 

In motor learning, feedback is studied as a means to regulate user behavior. Feedback can, for example, improve performance or technique in tasks that involve the human motor system~\citep{sigrist_augmented_2013}. The field views feedback as knowledge and has developed sub-concepts specific to motor learning strategies, like \textit{knowledge of performance} (feedback describing the technique or nature of the movement) and \textit{knowledge of results} (feedback reflecting the interaction outcome)~\citep{schmidt_motor_2018}. 
Compared to feedback terminology in educational contexts, knowledge of results is closely aligned with outcome feedback, while knowledge of performance aligns with process feedback. 
However, contrary to motor learning, educational contexts often seek to achieve deeper learning and higher reflection, focusing on cognitively demanding mental tasks (e.g. problem-solving, selection), that get executed through motor actions~\citep{Butler_Winne_1995}.
{\color{rv1} Functional feedback can also take the form of cues, which intend to provide users information and dissapear once they are no longer needed~\citep{Dillman_2018}. Like feedback notions in motor learning, cues can be provided at different levels of granularity from continuous (e.g. a color changing as a person moves~\citep{Karaosmanoglu_2024}) to discrete (a color changing once the person moved correctly). Previous work have for example studied how cues can help players understand if they performed correct body movements~\citep{Karaosmanoglu_2024}, and explored how to use cues to direct players' gaze~\citep{Lankes_2022}.}

The present study focuses on the motor learning goal of optimizing users' motor skills to perform 
continuous movements (like tracking an object or steering a car). These movements are characterized by relatively low cognitive involvement, with the nature of the movement itself being a key determinant of success~\citep{schmidt_motor_2018}. 
Motor learning refers to feedback dynamically driven by human performance as biofeedback~\citep{brennan_feedback_2020}, and is introduced into motion-controlled systems as augmented feedback, supplementing user's inherent bodily feedback while they perform, to benefit skill learning and performance~\citep{phillips_harnessing_2013}.
Games additionally employ dynamically driven feedback as a means to 
invoke curiosity, create sensual feedback, a more reactive environment, more variation, reward, and juiciness~\citep{Deterding_2015,Schell_2019,HicksFramework2018}. \textit{Fruit Ninja}\citep{halfbrick2010fruitninja}, \textit{Osu!}\citep{herbert2007osu}, and \textit{Beat Saber}\citep{beatgames2019beatsaber} continuously track players' input devices and use it to drive audiovisual effects dynamically as users play.
In \textit{Osu!}, a green trail of particles follows the player's cursor, which grows when players move the cursor at a higher speed. However, although the \textit{Osu!} trail rewards high cursor speed, the rewarded behavior does not align with what is needed to obtain better performance or technique. {\color{rv1} These types of feedback are often less explicit than cues, providing information through ambient audiovisual effects driven by the characteristics of user actions}. Game designers may consider feedback dynamics through, for example, ~\citeauthor{Schell_2019}'s lens of channels and dimensions, which demonstrates dividing games' interface into channels for game information~\citep{Schell_2019}. However, as \citeauthor{Schell_2019} admits: ``\textit{This is usually done partly by instinct, partly by experience, and mostly by trial and error} [..].''~\citep{Schell_2019}.

\subsection{Design Space of Feedback}\label{sec:dimensionsfeedback}
\begin{figure}[h]
      \centering
      \includegraphics[width=\textwidth]{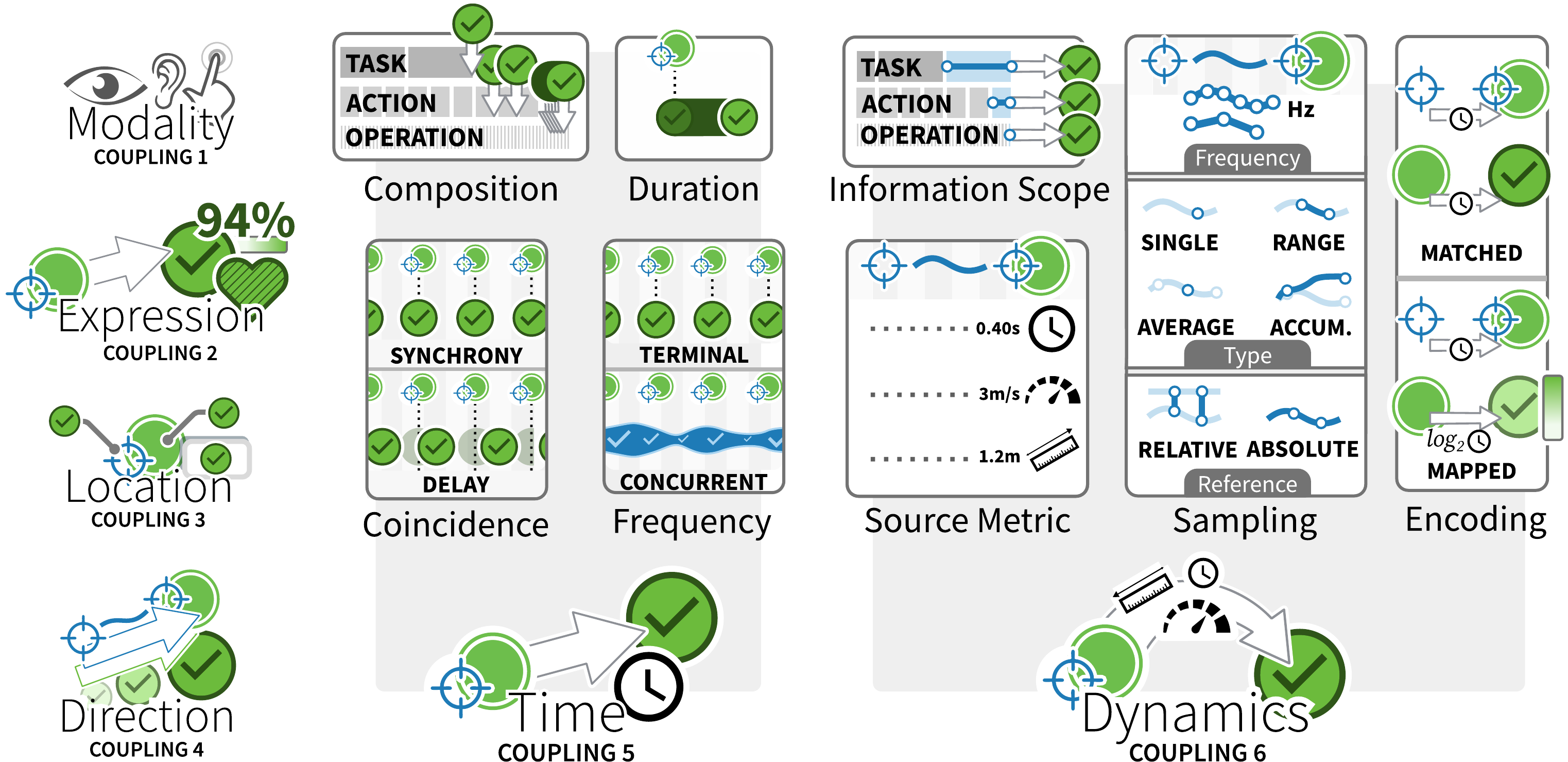}\caption{Using our study's pointing game as example, we created a conceptual design space, visually exemplifying \citeauthor{wensveen_interaction_2004}'s six couplings (C1-C6), including nine sub-dimensions of Time and Dynamics. 
      }
      \label{fig:couplings}
      \Description{Diagrammatic overview of Wensveen's couplings. Each coupling is displayed with title and a diagram depicting the coupling's meaning. The modality couplings shows an image of an eye, ear and finger to represent visual, auditory and haptic modalities. The expression coupling shows how blue cursor activating a target can translate into percentage, a heart or checkmark symbolic. The location coupling depicts several checkmark icons placed with respect to different objects or in a notification banner. The direction coupling depicts two arrows surrounded by a gradually moving cursor and gradually better feedback. The Time and Dynamic couplings both mostly occupies the figure, and lists their identified subproperties, visually represented using similar graphical depiction. The time coupling's coincidence property contrasts synchrony and delay. The frequency property contrasts terminal and concurrent feedback. The composition property contrasts task, action and operation aligned feedback. The dynamics coupling contain six properties. The conversion property contrasts matched and mapped conversion. Information scope contrasts information scoped to the task, action or operation-level. The measurement reference contrasts relative and absolute measurements. Finally, measurement type contrasts single, range, average and accummulative measurement types.}
\end{figure}

Although rarely used in game design contexts, \citeauthor{wensveen_interaction_2004}'s product design framework~\citep{wensveen_interaction_2004} offers a comprehensive starting point to unpack dimensions of feedback, illustrated in Fig.~\ref{fig:couplings}. 
\citeauthor{wensveen_interaction_2004}'s framework explain feedback design through six couplings:  Modality (e.g. audio/visual/haptic sensory channels), Expression (e.g. color, size, form factor), Location (e.g. spatial placement), Direction (e.g. alignment to the action's direction), Time (e.g. feedback's temporal occurence), and Dynamics (e.g. how information is mapped to appearance). 

Choice, combination and congruency of \textit{modality} have been extensively studied within HCI and motor learning~\citep{kim_benefits_2008, akamatsu_comparison_1995, Hooren_2020,Theil_2018,dix_role_2023,sigrist_augmented_2013}. For example, \citeauthor{kim_benefits_2008} demonstrated the performance benefits of congruent multi-modal feedback (repeating information across modalities in synchrony)~\citep{kim_benefits_2008}. 
Motor learning has studied forms of \textit{expression} as positive/negative \textit{framing}~\citep{schmidt_motor_2018}, natural or abstract representations (e.g. video feedback vs. numerical gauge), and granularity of representation (e.g. 5 shades of green versus a continuum of colour)~\citep{sigrist_augmented_2013}. Expression also represent game-centric notions like game feel~\citep{Swink_2014}, aesthetic appeal, and abundance (\textit{juiciness})~\citep{Hicks_2019}. 
The \textit{location} of feedback is commonly manipulated by games in superimposed heads-up displays~\citep{caroux_influence_2016}. It is for example common to show an avatar's health bar in the display corner, instead of next to the avatar itself. 
The feedback's \textit{direction} refers to the feedback's way of changing compared to the direction of the user's action~\citep{wensveen_interaction_2004}. For example, higher movement speed might yield better feedback (alignment), whereas longer movement time yields worse feedback (inverse alignment).

The \textit{time} coupling represents the alignment between user actions and the feedback~\citep{wensveen_interaction_2004}. Time has multiple facets, which we illustrate as four sub-dimensions in Fig.~\ref{fig:couplings}: Duration, Coincidence, Frequency and Composition. Temporal duration refers to the amount of time the feedback is shown (e.g. in seconds or milliseconds). Temporal coincidence (or \textit{timing}~\citep{Maresch_Donchin_2022}) refers to how feedback coincides with user interaction from synchronous (immediate) feedback to delayed feedback. In games, synchronous feedback in games ensures that players know what they need to know at a given moment~\citep{Schell_2019}, whereas delayed feedback (e.g. latency) can disrupt players' flow~\citep{Deterding_2015,dix_role_2023}. 
Temporal frequency represents how often users receive feedback in response to their actions. Terminal feedback (single discrete feedback instances at a frequency) and concurrent feedback (feedback shown continuously during the movement) are frequently compared in motor learning~\citep{sigrist_augmented_2013}.
Temporal composition represents the alignment of feedback to users' tasks and actions in games. 
\citeauthor{Hougaard_Knoche_2024}'s core task framework, for example, provides a process to deconstruct gameplay into a hierarchy composed of tasks and actions.
Composing feedback is the process of choosing when and where to provide it in the task hierarchy -- at the end of each task, each action, or continuously during subconscious operation~\citep{Hougaard_Knoche_2024}. 
For example, Fruit Ninja~\citep{halfbrick2010fruitninja} yields operation-level feedback (larger cutting trail) while the user cuts the fruit. Once the fruit is cut, the fruit becomes halved creating action-level feedback (fruit stains). The task ends when all fruits are cut, creating task-level feedback such as high score (number of fruits cut, and best fruit combo).

\subsection{Dynamic Feedback}
\citeauthor{wensveen_interaction_2004}'s last coupling, \textit{Dynamics},  refers to how the action maps to the system's reaction. \citeauthor{wensveen_interaction_2004} exemplifies such a mapping process with their scissor example: \textit{``The speed of the cutting action determines the speed of the incision being made.''}~\citep{wensveen_interaction_2004}. 
In digital games, dynamics refer to how players' performance is mapped to feedback by undergoing an encoding process. Performance can for example map directly to a change in color, or be augmented through a logarithmic conversion. 
Such encodings are also known as information-display mappings within HCI, although they are not specific to visual displays. 
Within motor learning, dynamic feedback serves to improve user skill and learning~\citep{sigrist_augmented_2013}. 
Design considerations include how the underlying performance measurements are chosen (source performance metric)~\citep{phillips_harnessing_2013}, the range of information used from the user interaction~\citep{Hougaard_Knoche_2024}, and how it is technically processed (e.g. numerical conversion, sampling method). 
We expand the dynamic coupling into four sub-dimensions in Fig.~\ref{fig:couplings}. The first dimension, \textit{Information Scope} refers to the context of information utilized in the dynamic feedback. \citeauthor{wensveen_interaction_2004}'s example of naturally mapped dynamic feedback only utilized information from the action instigating it~\citep{wensveen_interaction_2004}, but it could just as well utilize information from all so far performed actions or from a selected range of fine-granular operations performed within the finished action~\citep{Hougaard_Knoche_2024}. 
\textit{Source Metric} refers to the measured information source, such as the \textit{speed} of the cutting action in \citeauthor{wensveen_interaction_2004}'s scissor cutting example. Choosing a source performance metric is often task-dependent~\citep{phillips_harnessing_2013} -- for example, performance metric for real-time feedback in running disciplines were chosen by comparing the strength of evidence with running economy and injury~\citep{Hooren_2020}. Pointing tasks commonly employ movement time and target distance as performance metrics along with target size to calculate throughput from Fitts Law~\citep{Fitts_Peterson_1964}.
\textit{Sampling} refers to how information from a given source metric is sampled and stored prior to encoding, including the available measurement unit (sampling reference), and information granularity (sampling frequency and type).
Finally, \textit{encoding} refers to how the information is eventually translated into a specific type of feedback -- for example matched to a change in color.

Dynamic feedback has been indirectly studied in the context of exergames (games designed for physical exertion) and for novel input methods for games. For example, \citeauthor{Nacke_2011} studied how gaze, temperature, or galvanic skin response could be used as physiological source metric to control game elements, providing players dynamic feedback~\citep{Nacke_2011}. Their study focused on both direct and indirect continuous control via physiological input, influencing active elements (e.g. enemy targets) and passive elements (e.g. weather conditions)~\citep{Nacke_2011}. In both cases, feedback was concurrent with interactions and based on operation-level information, where feedback was synchronized in real-time to changes in users' physiological state. 
Data underwent downsampling, smoothing, and normalization as part of encoding of the physiological data~\citep{Nacke_2011}.
In a similar vein, \citeauthor{Ioannou_2019} explored dynamic feedback in a virtual reality (VR) run and jump exergame by amplifying players' inherent jump height~\citep{Ioannou_2019},
thereby augmenting their inherent feedback. They measured jump height from the ground floor (source metric) for each individual jump action (information scope) and used it as multiplication factor for encoding the resulting feedback. 
Both studies found experiential benefits of incorporating dynamic, be it as a dramatic device~\citep{Nacke_2011}, or to increase intrinsic motivation and perceived competence~\citep{Ioannou_2019}. However, it remains unclear what processes are involved when players regulate their performance with dynamic feedback, and what experiential effects design choices like choice of performance metric and timing have, particularly for more common tasks like pointing.

\subsection{Research Gap}\label{sec:scope}
Players may find dynamic feedback rewarding in games, but it is unclear to designers when to deploy feedback schemes that help players improve {\color{rv1}and where they can} exercise artistic freedom without causing performance interference. 
Previous work studied effects of juicy feedback~\citep{Hicks_2019,Kao_2024}, dynamic feedback based on physiological input~\citep{Nacke_2011} and in motor learning contexts~\citep{Hooren_2020,sigrist_augmented_2013}, but to create a better foundation for the dynamic feedback design process, the relationship between feedback, performance metric, performance, and users' sense of correspondence needs to be clarified:

\begin{itemize}
\item[RQ1:] How does metric-based dynamic feedback improve or penalize people's performance? For example, what improvements or penalties occur when players utilize dynamic feedback of their own peak speed during pointing?
\item[RQ2:] What cognitive factors exist in players' self-regulation loop and do players feel dynamic feedback affects them?
\end{itemize}

To address this gap, we 1) conceptualized a model of performance regulation in a simple pointing game, and 2) utilized the pointing game to systematically compare changes in players' performance and perception of dynamic feedback, driven by three performance metrics, promoting three distinct behaviors: lower completion time, higher peak speed, or straighter movement. 
By providing players' knowledge and first-hand experience of how each metric worked, we hoped to create optimal conditions for measuring how players' self-regulated with dynamic feedback.

\section{Conceptual Model}\label{sec:conceptual-model}
Understanding dynamic feedback's effect upon performance self-regulation requires knowledge of the cognitive processes involved in the underlying interaction loop between user and system.
Human-computer interaction has commonly been conceptualized based on \citeauthor{Norman_1988}'s Stages of Action in which feedback partially reveals the system's state to the user who interprets it to determine their next course of action~\citep{Norman_1988}. 
Games similarly consist of loops, where a simple task is repeated over and over -- also called `\textit{low-level action-feedback loop}'~\citep{Kao_2024} or \textit{a mastery feedback loop}~\citep{Rigby_Ryan_2011}. Unlike progression-based game loops in which skills build on each other to form complex skill chains (e.g. ~\citep{Horn_2017}), low-level action-feedback loops are characterized by high repetition and focus on enveloping players in \textit{success-dependent feedback}~\citep{Kao_2024}, which reward players the more they master the task~\citep{Rigby_Ryan_2011}. 

An example of a simple game loop, is the target-to-target pointing task, which consists of targets appearing one after another, resembling \textit{immediate-response tasks} from information processing theory in cognitive psychology~\citep{Newell_1994}. The immediate-response task is a subtype of the \textit{speeded response task}, which aim to ``\textit{produce a simple response dependent in some well-specified direct way upon the presentation of a stimulus.}''~\citep[p.261]{Newell_1994} In target-to-target pointing tasks, the simple response consists of a quick movement from one target to the next. Looped tasks are said to be data-dependent -- the output of the previous action becomes the input of the next.
The loop between user and a system is also known as micro-level monitoring in self-regulated learning, where users use self-observation and system feedback to determine whether they are progressing appropriately to regulate their effort and affect (e.g. frustration or motivation)~\citep{Panadero_2017}.

\citeauthor{Newell_1994} categorizes immediate-response tasks into eight stages: perception (sensing the environment including feedback), encoding (parsing the information), attending (focusing on relevant details), comprehending (analyzing the information), setting the task (deciding what to do), intending it (determining when to act), decoding it (converting to commands), and finally movement (executing the action)~\citep{Newell_1994}. 
If a stage fails (e.g. failing to perceive the feedback), it hinders the creation of a functional pathway from perception to movement in \citeauthor{Newell_1994}'s model.
Similarly in game contexts, \citeauthor{Kao_2024} emphasized that feedback must be both legible and differentiated for fun and motivation to arise: ``[..] \textit{even at low-level action-feedback loops, motivation and enjoyment arise from reducing uncertainty over action success, which depends on \textit{legible} and \textit{differentiated} amplified feedback.}''~\citep{Kao_2024} 
Legibility refers to the \textit{causal action-feedback link} (knowing I created an outcome) while differentiation expresses how well an action was executed.

We conceptualized a self-regulation model for pointing tasks (Fig.~\ref{fig:conceptualspace}), which encompass the system, world and mind, based upon \citeauthor{Norman_1988}'s Stages of Action and \citeauthor{Newell_1994}'s scheme for immediate-response tasks. 
We drew upon existing scales to develop and iterated upon the measures through pilot tests in parallel with our study's apparatus and procedure, further detailed in Supplementary Material 1. 

\begin{figure}[h]
      \centering
      \includegraphics[width=\textwidth]{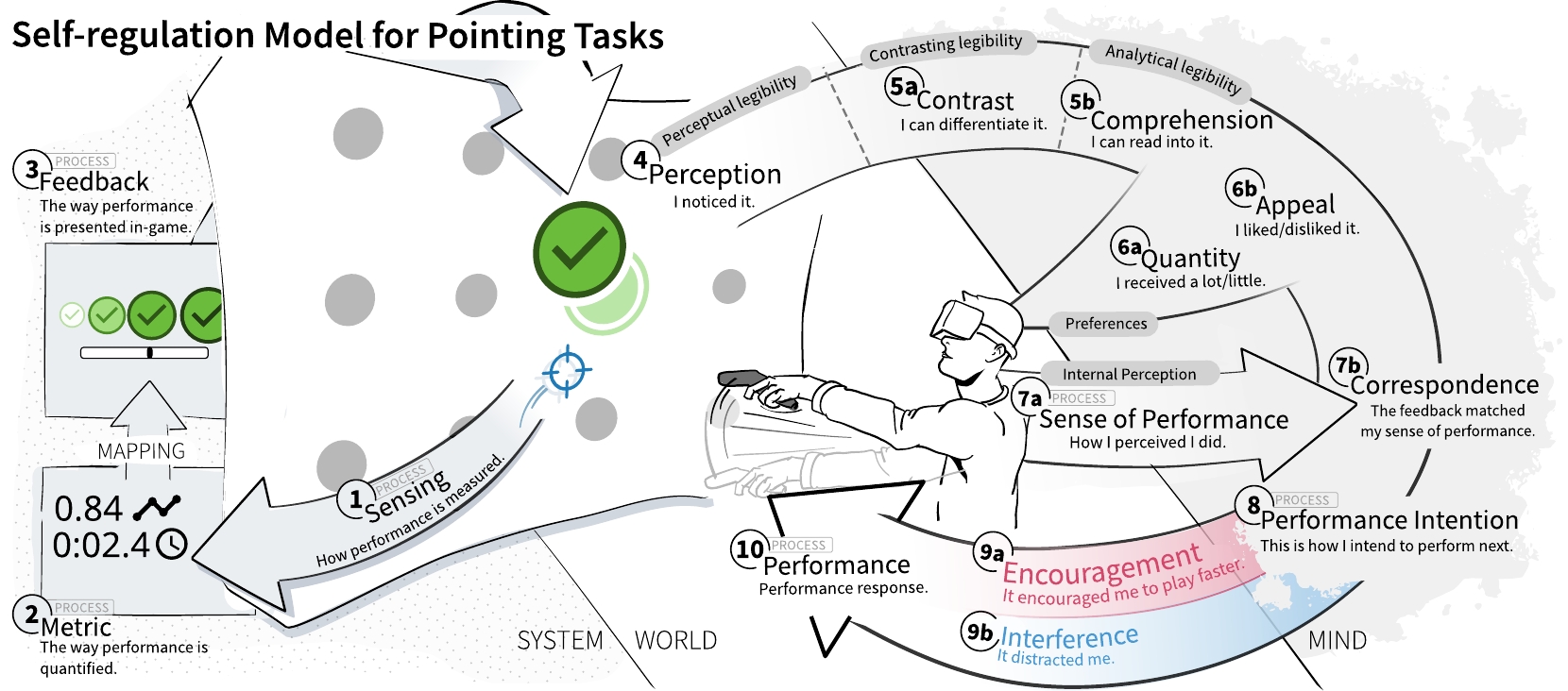}
      \caption{How a VR pointing game's feedback (e.g. checkmarks) regulates a player's movement performance. The player move their arm to move a crosshair to the target. The game senses that motion (1), quantifies it (2), and presents \textit{feedback} (3) perceived by the player (4-5). They assess the feedback (6) and match it to their sense of performance (7), which may affect subsequent performance (8-10). {\color{rv1}The model is a conceptual framework developed by the authors, based upon \citeauthor{Norman_1988}'s Stages of Action and \citeauthor{Newell_1994}'s scheme for immediate-response tasks.}}
      \label{fig:conceptualspace}
      \Description{Self-regulation Model for Pointing Tasks. A drawn person, seen from the side, is wearing a head-mounted display and moves a controller around to point a blue crosshair at a green target with their left arm. The person is surrounded by a virtual world containing a large wall of targets, annotated by a loop of arrows and boxes with text. An arrow titled sensing, points from the moving crosshair to a textbox tiled metric, described as the way performance is measured and quantified. The metric textbox points to a textbox behind the wall titled feedback, described as the way performance is presented in-game. A large arrow points from a checkmark on the wall into the person's mind. Different stages of legibility are shown along the way. In the perceptual legibility stage, perception is described as noticing something. In the contrasting legibility stage, Contrast is described as I sensed the dynamic. Analytical legibility is at the first crossing with user preferences, including subjective perceptions of quantity and appeal. Finally, the arrow leads to correspondence, described as the feedback matched my sense of performance. A different arrow points from the user to correspondence called internal perception, where the sense of performance describes how the user perceived they did. The arrow further leads to Comprehension, which constitutes the ease with which users can read what they see. From here the arrow ends at performance intention, which is the beginning of a new arrow pointing back into the virtual world. It lists three concepts in a red, blue and white background color. The red concept is titled `encouragement' and subtitled, which encouraged me to play faster. The blue concept is titled interference and subtitled it distracted me. The white concept is titled performance and is shown at the end of the arrow, which points at the person's moving arm.}
\end{figure}

\subsubsection{System Stages 1-3}
Represents the underlying algorithms that map user performance to feedback. 
User performance is measured in real-time by obtaining data such as movement position at a given sampling rate from the VR headset and sensors in the \textit{sensing} stage (1). A performance metric such as speed is derived from the sensor data, which is then normalized to an abstract form, like percentage in the \textit{metric} stage (2).
Finally, the abstract representation is translated into feedback with specific intensity and occurrence in time in the \textit{feedback} stage (3).

\subsubsection{Legibility Stages 4-5}
Based upon \citeauthor{Kao_2024}'s notion of legibility, perceptual legibility, contrasting legibility, and analytical legibility are sub-concepts, derived from perception, encoding, attention and comprehension stages in immediate-response tasks. Users' notice the feedback in the \textit{perception} stage (4), influenced by design features in the system that help users see presented information (the sensory affordances)~\citep{Hartson_2003}. 
The \textit{contrast} stage (5a) to follow, captures users' ability to recognize changes to the feedback's appearance over time. 
Then in the \textit{comprehension} stage (5b), users mentally connect their action to the system's reaction to make causal assessment, which is influenced by design features that help users facilitate knowledge and thought (cognitive affordances)~\citep{Hartson_2003}. 
Comprehension is similar to PXI's \textit{Progress Feedback} sub-measure ``\textit{I could easily assess how I was performing in the game}'' and to \textit{Feedback Ambiguity} in \citeauthor{Hicks_2019}'s framework for juicy game design (``\textit{Can information be connected to actions and only interpreted in one way?}'')~\citep{HicksFramework2018}.

\subsubsection{Preference Stage 6} 
Where users' expectations and subjective feedback preferences are superimposed onto the evaluation process~\citep{Kao_2024}.
The \textit{quantity} stage (6a) captures how much feedback players feel they receive -- if feedback is too excessive,
it can unintentionally reduce enjoyment and playtime, highlighting the importance of appropriate feedback amplification~\citep{Kao_2024}.
Perceptual \textit{appeal} (6b) refers to 'consummatory likeness'~\citep{Kao_2024} -- the interest and enjoyment players derive from the feedback.

\subsubsection{Internal Perception Stage 7}
Refers to the internal information users store in memory of the action they performed -- users' \textit{sense of performance} (7a). Motor learning conceptualized that as \textit{inherent feedback} -- the memory of how the movement felt, sounded and looked, which is accessible to people even before their actions are completed, enabling self-detection of error (or \textit{subjective reinforcement}~\citep{schmidt_motor_2018}).
We introduced the correspondence stage (7b) to record how users sensed their internal performance matched externally perceived feedback. It relates to \textit{unambiguous feedback} from the Flow State Scale~\citep{Jackson_Marsh_1996} (``\textit{I had a good idea while I was performing about how well I was doing.}'').  

\subsubsection{Performance Intention (8-10)}
Represents the closest stages to motor system actuation and extends the \textit{intend} and \textit{move} stages in immediate-response tasks. It consists of a \textit{performance intention}, which is decoded into a \textit{performance response}, which is affected by the extent which feedback was \textit{encouraging} or \textit{distracting}. Encouragement and distraction 
are somewhat inverse to the \textit{positive effectance} measure (users' ``\textit{perceived influence on the game world}'')~\citep{Klimmt_2007}, which measure how effectively motor work yielded more feedback -- by contrast, distraction represents the perceived cost of feedback, while encouragement represents the perceived gain of feedback.

\section{Study}\label{sec:feedback-study}
We designed a within-participant experiment to study 24 voluntary participants who played a virtual reality (VR) pointing game for 30-40 minutes {\color{rv1} with their right hand}.
{\color{rv1} 
We utilized pointing as a platform to study how well participants could adapt their performance, following three different source metrics as main study's main variable -- the primary component to create dynamic feedback in Wensveen's design space in Fig.~\ref{fig:couplings}.
To visualize the metrics, we created three feedback designs. We used three different compositions  (Time coupling in Fig.~\ref{fig:couplings}) as the design anchor to make each feedback appear distinct and let the composition become an anchor for determining the appropriate expression and location. 
Performance and all cognitive factors from the self-regulation model in Fig.~\ref{fig:conceptualspace} were utilized as dependent variables, to measure players penalties or improvements, including perception, contrast, comprehension, quantity, appeal, correspondence, encouragement and interference. 
}

Using an iterative feedback design process, we aimed to effectively unveil a broad spectrum of potential penalties and improvements in the conceptual model's performance, legibility, preference, and performance stages in Fig.~\ref{sec:conceptual-model}. The study employed differently timed dynamic feedback that provided different levels of exposure to the underlying metric. They leveraged metaphors from well known phenomenons -- pointer trails used in most desktop accessibility settings to increase cursor visibility or recognizable and affirmative checkmark symbols.
The feedback followed a fixed order, from high exposure to low exposure, ensuring that players' had high familiarity with each metric, even with low access to feedback. The study made measurements across all three sessions to identify overall improvements or costs to performance.

The study was approved by the ethical review board of the lead author's institution (REB \#2023-505-00051). 
The study examined how dynamic feedback influenced players' performance and experience using three metrics: distance, time, and peak velocity. 
\subsection{Apparatus: VR Pointing Game}

\begin{figure}[h]
      \centering
      \includegraphics[width=\textwidth]{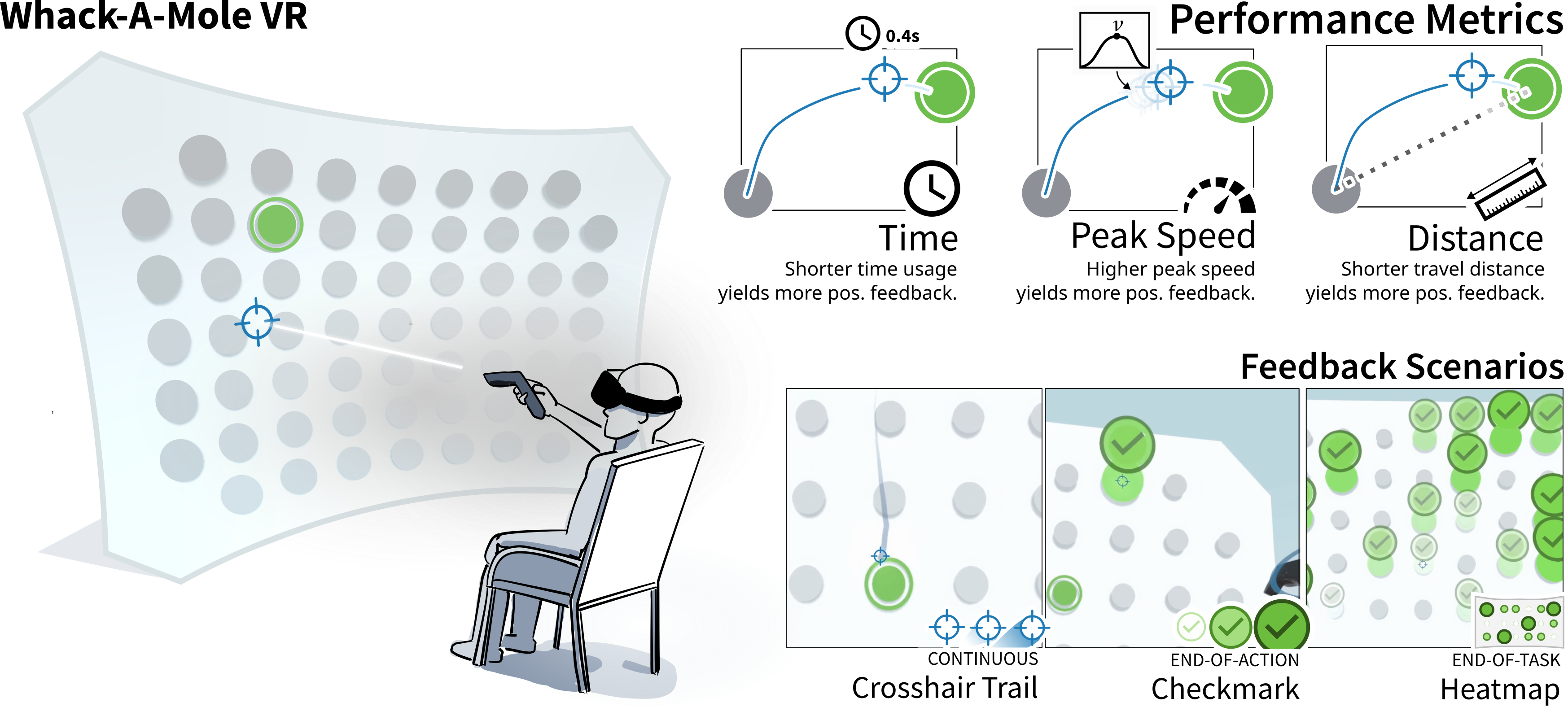}
      \caption{Whack-A-Mole VR consists of a large $6 \times 3$ meter wall, and a cursor controlled by the player from their spot (left). Three {\color{Highlight} positive }feedback types were tested (upper right) and were {\color{Highlight} moderated by} three different performance metrics (lower right).}
      \label{fig:whack-vr-overview}
      \Description{Depiction of Whack-a-mole VR, alongside the study's feedback and metrics. On the left side, titled Whack-A-Mole VR, a floating black game controller is pointed at a tall rectangular grey wall, its exact pointing location determined by a blue crosshair. The grey wall contains five rows and nine columns of circular gray targets, except for a single green target in the upper third column. The upper right side, titled feedback, shows three square thumbnails. The leftmost thumbnail, titled continuous crosshair trail, depicts a crosshair approaching a green target while leaving behind a faded blue trail. The middle thumbnail, titled end-of-action checkmark, depicts a green circular checkmark with a heavy green outline hovering above a target. The rightmost thumbnail, titled end-of-task heatmap, depicts a dozen green targets with circular checkmarks above them at varying green intensities. The lower right side, titled Metrics, shows three diagrams depicting performance metrics. The leftmost diagram is titled Time and described as: Shorter time usage yields better feedback. The diagram depicts a crosshair's trajectory, with a black stylized clock shown above it. The middle diagram is titled Peak Speed and described as: Higher peak speed yields better feedback. The diagram depicts a crosshair with motion blur with an overlaid diagram depicting a velocity curve. The rightmost diagram is titled Distance and described as: Shorter travel distance yields better feedback. The diagram depicts a crosshair's trajectory as a line and a shorter straight Euclidean trajectory as a dashed line.}
\end{figure}

The study's pointing task consisted of playing a Whack-A-Mole VR game~\citep{Hougaard_Knoche_Brunner_Evald_2022} on a HTC Vive head-mounted display while sitting down, shown in Fig.~\ref{fig:whack-vr-overview}. 
Whack-A-Mole VR~\citep{hougaard2019whack} was chosen because it was designed for continuous performance data collection for rehabilitation setting at high granularity; it provided calibration procedures, a simple distraction-free VR environment, and configurable activation patterns to make participant data comparable~\citep{Hougaard_Knoche_Brunner_Evald_2022}.

In Whack-A-Mole, players controlled a blue crosshair to hit round green targets on a 2D horizontal $5m \times 9m$ grid shown in front of them in VR. Targets were activated by hovering over a target for 400ms -- no button pressing was necessary to avoid the Heisenberg effect (parasitic motion from the motion-tracked controller when pressing its button down)~\citep{bowman_using_2001}. Every target was active for 5 seconds and appeared immediately after each other (target-to-target). The crosshair moved in response to horizontal and vertical controller movements, unaffected by changes to controller angle -- players pointed by moving their controller around within a small invisible $35cm \times 25cm$ motor space, calibrated to be within reach from players' seated position (e.g. as illustrated in~\citep{Hougaard_2022}). 
We incorporated three performance metrics (\textit{completion time}, \textit{peak speed}, and \textit{travel distance}) which we tested with distinctly different feedback in three sessions:
\begin{enumerate}
    \item[(S1)] \textbf{Trail.} Continuous operation-level feedback which consisted of a trail, simulating a natural phenomenon observable on fast-moving objects in the real world, to maintain a natural expression and location coupling to the user's cursor movements.
    The blue trail matched the cursor's color, and the trail's length and opacity were moderated by performance (from 0  to 9m length and 0-100\% opacity). The trail emitted a continuous sound, 
    whose pitch from 195hz to 320hz dynamically varied with users' performance.
    Users received continuous haptic vibration in short 6ms pulses with variable amplitude, moderated by performance.
    \item[(S2)] \textbf{Checkmark.} The player hovers over each green target to hit them and receives a green checkmark when completing an action. The green circular checkmark is a well-known acknowledgement symbol used in game and non-game contexts and was co-located with the target to maintain a natural location coupling. 
    The checkmarks scaled 100-150\% of their baseline size and changed color from pale to saturated green, moderated by user performance. An auditory ``pling'' sound changed to brighter pitches between 514 Hz and 817Hz continuously across 8 semitones. Players received a short 50ms haptic vibration with variable amplitude.     \item[(S3)] \textbf{Heatmap.} 
    After all targets were hit, a heatmap appeared on the wall with a 3-second introductory animation, where targets each emit a checkmark by hitting order. With each checkmark, the player's controller vibrated shortly for 100ms and they heard a brief ``pling'' sound at variable pitch forming a 3-second melody, which represented player performance. To give players ample time to see the heatmap, it was visible for 10 seconds.
\end{enumerate}

All feedback in S1-S3 was multimodal, via sound from headphones, visuals in VR, and haptic vibration feedback in the hand controller. Expression-wise, the feedback changed size, color saturation, vibration intensity, vibration frequency, audio pitch, and audio volume. All feedback was located in proximity to the element of interaction and featured similar directionality -- receiving \textit{more} feedback always meant objectively performing \textit{better}.

\subsubsection{Combining Metric and Feedback}
We designed the three performance metrics \textit{completion time}, \textit{peak speed} and \textit{travel distance} shown in Fig.~\ref{fig:whack-vr-overview}, bottom. The metrics were not designed as "fair" or ideal models of human pointing performance, for example as derived from Fitts Law~\citep{Fitts_Peterson_1964} and studied by kinematic literature~\citep{meyer_speedaccuracy_1990,Soukoreff_MacKenzie_2004}). Instead, each metric focused on an isolated aspect of human performance, making it possible to measure whether or not participants adapted their performance to the metric.
The metrics's information scope matched the feedback's temporal composition - for example, action feedback only reflected performance information from the specific action. The metrics were measured as single samples of data with at 60Hz sampling frequency, and initially tested in textual form (e.g. ``80\%'', ``2.3m'', ``4.1s'') to ensure the correctness of measured performance. 
The metrics were implemented for each feedback type, like described in Fig.~\ref{fig:patterns-study}.

\begin{figure}[h]
\centering
\includegraphics[width=0.9\textwidth]{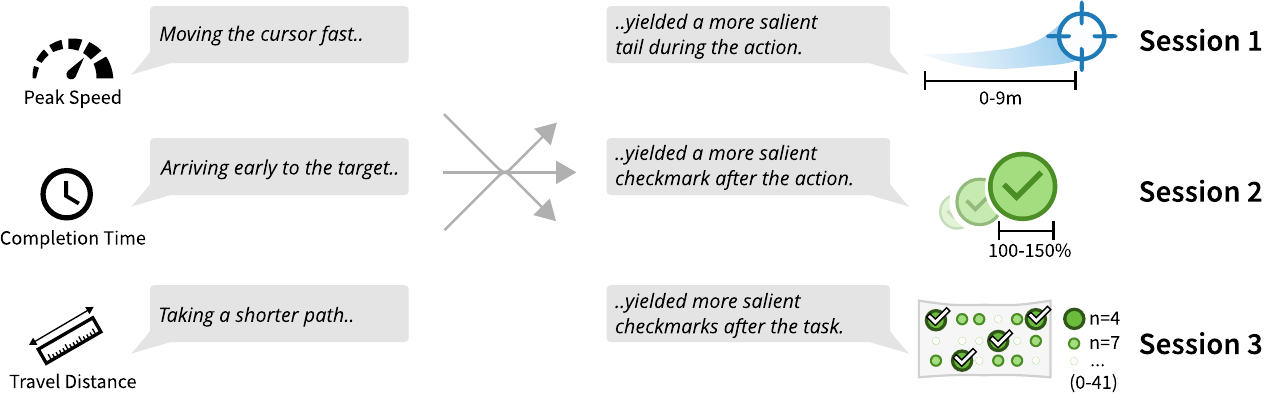}
\caption{Overview of the how the three performance metrics (peak speed, completion time, travel distance) combined with each feedback type, {\color{rv1} which changed its saliency (e.g. how much it was standing out through size, color, and opacity)}}
\label{fig:patterns-study}
\Description{A diagram consisting of performance metric on the left side and trail, checkmark and heatmap feedback, featuring different timing in three separate sessions. The diagram shows how to combine each metric with feedback using dialog metaphor. Peak speed's dialog is "moving the cursor fast", completion time is "arriving early to the target" and travel distance is "taking a shorter path". Trail feedback's dialog is "..yielded a more slient tail during the action". Checkmark feedback's dialog is "..yielded a more salient checkmark after the action". Heatmap feedback's dialog is "..yielded more salient checkmarks after the task.".}
\end{figure}

The time metric measured the time passed in seconds from action start to either the current cursor position (crosshair trail) or to the end of the action (checkmarks) -- it rewarded players when they optimized their time usage and punished actions with high time usage. 
The peak velocity metric measured the current velocity at the cursor position (crosshair trail) or the highest velocity that happened during the action (checkmarks) --
by design rewarding swift movements, and punishing slow movements.
The distance metric measured the accumulated distance (measured in nearest whole cm) divided by the shortest Euclidean distance to either the current cursor position (crosshair trail) or to the end of the action (checkmarks) -- by design it rewarded straighter movements (\textit{trajectory straightness}) while punishing behavior that led to more accumulated travel like hand shaking or overshooting.

\subsubsection{Adaptive Feedback System}
To map metric to feedback, we implemented a feedback system, which measured and assessed users' game performance in real-time, relative to observations of minimum and maximum performance peaks. {\color{rv1} the feedback system was calibrated to account for individual player performances, by determining floor and ceiling through averaging the five most recent worst and best performances.} Using an average instead of a single observation or an absolute value increased the systems' robustness against coincidental minimum/maximum performance outliers stemming from noise, learning effects or fatigue. Additionally, 1\% of participants' peak performance ranges were continuously subtracted from the minimum and maximum peaks every second, ensuring that the system continually obtained new minimum and maximum peak values over time. When the feedback system had a low number of observations (i.e. $<$ 3 hits) it emphasized rewarding users by design, while capably providing more nuanced and accurate rewards when it had more observations to estimate user performance from. To provide a fair comparison between conditions, the feedback system reset with every condition. {\color{rv1} During play the feedback system's floor, ceiling and assessments were tracked, and to help ensure they matched users' performance.}

\subsection{Study Design}\label{sec:study-design}
Following the objective of understanding how people's adapted to three performance metrics, we designed a study constituting three counterbalanced sessions shown in in Fig.~\ref{fig:patterns-study}. Each counterbalanced session followed a Latin square, so each participant had a different performance metric order (e.g. Time $\rightarrow$ Speed $\rightarrow$ Distance). We captured performance measurements from participant actions, as they hit each target on the Whack-A-Mole wall.  
We used perceptually random performance balanced activation patterns, that maintained high repeatability across conditions (detailed in Supplementary Material 2). Completing the experiment required each participant to move their limb 44.9 meters Euclidean distance, verified through data logging.
{\color{rv1}
We chose a fixed order for the feedback types, that maximized participants' exposure to the underlying source metrics, by letting everyone experience continuous operation-level feedback first (most exposure). This way, participants were given more optimal conditions to self-regulate during short exposure to the non-continuous feedback that followed (end-of-action checkmarks, and end-of-task heatmap). This desirable learning effect was became the cornerstone of the study's final within-subject design which economically balanced the total number of playthroughs (9 per participant) with the overall study length (30-40 minutes).

}
Numerous pilot tests refined the study's apparatus (feedback system, types, metrics), and experimental design (conditions, control variables), to ensure consistent performance metric implementation across feedback types with adaption to all ranges of human performance. The feedback's auditory, visual and haptic design was improved by obtaining impressions from pilot test subjects, consisting of game design students and two persons working in the game industry. 

\begin{figure}[ht]
\centering
\includegraphics[width=\textwidth]{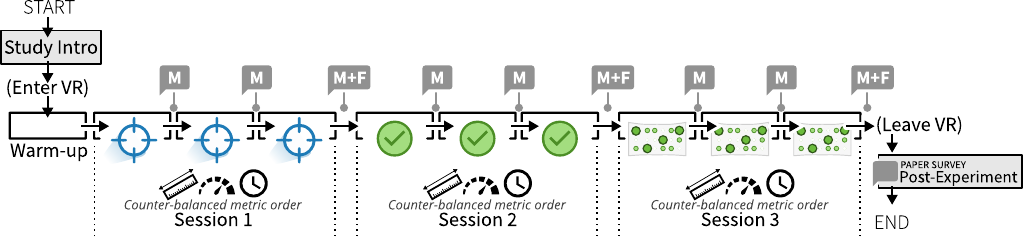}
\caption{Study timeline, showing the $3\times3$ design featuring three sessions with three counter-balanced conditions in each, and the occurrence of metric surveys ('M'), feedback surveys ('F') and the post-experimental survey which included a short interview.}
\label{fig:patterns-study}
\Description{Diagram showing the timeline of the study from start to end, divided into stages by outlined rectangles. The first stage is the study intro. An arrow points to an intermediate stage named "Enter VR". The stage afterwards is "Warm-up". Three sessions are then shown, with three stages in each. In session 1, evey stage features trail feedback. In session 2, every stage features checkmark feedback. In stage 3, every stage features heatmap feedback. Between stages dialogs with "M" are shown, while dialogs with "M+F" are shown between sessions. After all three sessions an arrow points to an intermediate stage called "Leave VR". The arrow finally points to "Post-Experiment Paper Survey" before reaching the end.}
\end{figure}

\subsection{Participants}
Fifteen male and nine female participants between 22-41 years old ($M$=$27.2$, $SD$=$3.9$) volunteered their time. 
Almost all considered themselves experienced at playing video games (22/24), and the majority (19/24) had previous experience with VR. Half of the participants (14/24) wore glasses and were allowed to take them off if they could clearly read the instruction text without them. None of the participants reported having previously played Whack-A-Mole VR before.

\subsection{Procedure}
At the start of the study, participants filled out their consent form, their gender, and age. Participants were informed that we measured how they experienced variations of a virtual reality game and that their personal data would be anonymously processed as per the ethical and data law requirements. 
The facilitator asked participants verbally about their previous VR and gaming experience and expectations of the experiment.
The facilitator showed them images of Whack-A-Mole VR and verbally instructed how to play, emphasizing that they were to perform their best.
Participants were introduced to the different feedback and performance metrics, and could ask questions if in doubt.
They then donned a VR headset and were calibrated to remain seated in the virtual scene's center.
If participants wore glasses, they either wore them inside the headset or took them off based on preference and whether the instruction text inside the HMD was legible.
A short warmup with binary audio-visual feedback confirming target hits lasted for 1 minute (green target disappears with an unvarying ``pling'' sound). They then played the first session with trail feedback reacting to participants' speed, distance, or time. Before each condition, the facilitator announced the upcoming performance metric. They had to hit 20 targets with each metric and were afterwards asked to rate their sense of correspondence (Fig.~\ref{fig:conceptualspace} ,7b) as shown in Supplementary Material 1. When the third and final metric was rated for trail feedback, they rated the feedback's quantity (Fig.~\ref{fig:conceptualspace}, 6a) and appeal (6b). They then proceeded to three conditions with checkmark-based and, finally, heatmap-based feedback where the procedure was repeated.
After completing all conditions, the participants filled out the post-study questionnaire (Supplementary Material 1) regarding their perception (Fig.~\ref{fig:conceptualspace}, 4), Contrast (5), encouragement (9a), and interference (9b). Finally, the participants were briefly interviewed about their experience and reported any felt discomfort, e.g. (cyber-)sickness or fatigue.

\subsection{Data Analysis}\label{sec:data-analysis}
Table~\ref{tab:measurement-overview} summarizes all quantitative-, survey-, and control measures. Participant movements were continuously recorded with 68Hz mean sampling frequency (30-95Hz, $SD$=$0.008$).
Data was imported into R Studio for analysis. Movements from each condition were split into individual actions based on target hits, which marked one action's end and the next action's start. 
We interpolated samples to time-equidistant values at 100Hz (10ms per sample) and calculated speed and position at each point and smoothed speed according to a degree 2 polynomial, following \citeauthor{Mandryk_Lough_2011}'s suggested approach~\citep{Mandryk_Lough_2011}. We calculated Fitts' ID to quantify the challenge of each pointing task independently from target-to-target distance and target width. Our quantitative performance measures in Table~\ref{tab:measurement-overview} were inspired by \citeauthor{Mandryk_Lough_2011}.
Movement time, peak speed, and straightness map directly to time, peak speed and distance metrics, which we visualized as bar charts following~\citeauthor{Mandryk_Lough_2011}.
{\color{rv1}
Quantitative measurements were tested for normality using a Shapiro-Wilk test and for sphericity using a Bartlett's test. Except for Throughput, all quantitative measurements were transformed prior to ANOVA analysis to follow a normal distribution. The appropriate transformation was determined through Box-Cox analysis, maximizing the log-likelihood function (See Supplementary Material 2). 
The quantitative measurements were analyzed using 3x3 repeated measures ANOVA with a Huynh-Feldt correction applied to p-values when sphericity assumptions were violated. Bonferroni correction with $\alpha$=$0.05$ was applied when making pairwise comparisons. Effect sizes were calculated using generalized eta square for RM-ANOVA and Kendall's W for Friedman tests.
Learning effects were estimated using visual trend line analysis and assessed within each feedback block, to estimate its effect separately from the fixed feedback ordering. Self-report measurements did not follow normality and were analyzed using a non-parametric Friedman test.
}

\begin{table}[th]
\renewcommand{\arraystretch}{1.1}
\centering
\resizebox{\linewidth}{!}{
\begin{tabular}{p{3.6cm}p{6.9cm}p{6.9cm}}
\toprule & 
 Description
 & Example/Question \\
\toprule
\textbf{Quantitative Measures} & (* Following \citeauthor{Mandryk_Lough_2011}~\citep{Mandryk_Lough_2011}) & \\
Arm Travel (meter) & How much distance did the users' arms travel while exercising?
 & Fx. 12.2 meters \\
Movement Time* (ms)  & How quickly did users perform their movement? & Fx. 1300 milliseconds  \newline (corresponds to time metric) \\
Straightness (0-1) & How straight did users perform their movement? (compared to ideal trajectory) & Fx. 0.83 (83\% straight)  \newline (correspond to distance metric) \\
Peak Speed* (m/s)  & What was the highest achieved speed of all actions?
 & Fx. 3 meters per second  \newline (corresponds to peak speed metric) \\
Time to Peak Speed* (ms)  & How long time did it take users to achieve the highest speed?
 & Fx. 532 milliseconds \\
Peak Speed to Target* (\%)  & How big a \% of the complete movement time was spent finishing the task after achieving peak speed?
 & Fx. 56\% \\
Fitts ID  & The measured average index of difficulty across the condition (designed to be equal across conditions).
 & Fx. 2.75 \\
Throughput (bits/s)  & Target distance- and width-independent measure of performance per second (index of performance~\citep{Fitts_Peterson_1964}).
 & Fx. 2.21 bits per second. \\
 \toprule
 \end{tabular}}
\caption{Overview of quantified game measurements in the study.}
\label{tab:measurement-overview}
\end{table}

Table~\ref{tab:experiential-survey} lists all experiential measures, which were based upon the conceptual model in Section~\ref{sec:conceptual-model}. Likert scale measures were visualized as violin plots, showing mean values, distribution and data points per participant. Raw performance data points were aggregated to means per participant, and Likert scale data points reflected each participant's rating.
Qualitative data was collected via note-taking and was analyzed thematically using an open coding analysis approach~\citep{Taylor_Bogdan_DeVault_2015}. We categorized responses into themes inductively without external review, and compared them to our own noted observations of participant reactions.

\begin{table}[th]
\renewcommand{\arraystretch}{1.1}
\centering
\resizebox{\linewidth}{!}{
\begin{tabular}{p{3.6cm}p{6.9cm}p{6.9cm}}
\toprule & 
 Description
 & Example/Question \\
\toprule
\multicolumn{2}{l}{ \textbf{Metric Survey} } & \\
Correspondence & How clearly the feedback corresponded to their own sense of performance. & ``\textit{With this algorithm, the feedback clearly corresponded to whether I was fast or slow.}'' \\
\multicolumn{2}{l}{ \textbf{ Feedback Survey } } & \\
Quantity & How much feedback users felt they received. & ``\textit{How much feedback did you get?}'' \\
Appeal & How much the feedback appealed to them. & ``\textit{Overall, how good did the feedback feel?}'' \\ 
\multicolumn{2}{l}{  \textbf{ Post-Experimental Survey } } & \\ 
Contrast & How well users were able to sense the difference between the algorithms. & ``\textit{With the [blue trail/checkmark/heatmap] feedback, I sensed the difference between the three algorithms.}'' \\
Perception & How much did users notice the feedback. & ``\textit{Overall, how much did you notice the [blue trail/checkmark/heatmap] feedback?}'' \\
Encouragement & How much the feedback encouraged players to perform. & "\textit{The [blue trail/checkmark/heatmap] feedback encouraged me to play faster.}" \\
Distraction & How much users found feedback distracting. & ``\textit{How much did you feel that the [blue trail/checkmark/heatmap] feedback distracted you?}'' \\
\leavevmode\color{Highlight} Comprehension  & \leavevmode\color{Highlight} How well users could use the feedback to assess their performance. &  \leavevmode\color{Highlight} "\textit{With the [blue trail/checkmark/heatmap] feedback I could easily assess how well I was performing.}" \\
\leavevmode\color{Highlight} Appeal  & \leavevmode\color{Highlight} How much the feedback appealed to them. & \leavevmode\color{Highlight} "\textit{Overall, how did the [blue trail/checkmark/heatmap] feedback make you feel?}" \\
Feedback Preference & Which of the three feedback types people preferred & ``\textit{Please indicate which feedback type you preferred:
} [Blue trail] [Checkmark] [Heatmap]'' \\
Subjective  & Which of the three feedback types people preferred & ``\textit{Please indicate which feedback type you preferred:
} [Blue trail] [Checkmark] [Heatmap]'' \\
\toprule
\end{tabular}}
\caption{Overview of each self-report measurement used in the study.}
\label{tab:experiential-survey}
\end{table}

\clearpage

\section{Results}\label{sec:results}
The participants (N=24) took on average 16~min and 27~s ($SD$=$3.13 min$) to hit 190 targets in total by moving their right arm around 50.5~m on average (5.6~m above required Euclidean distance). Table~\ref{table:study-condition-means} lists averages and standard deviation of measurements, categorized by metric and averaged across three sessions. 
{{\color{rv1} We report our quantitative results with means ($M$), standard deviation ($SD$), effect size ($\eta^2$) p-values ($p$), inter-quartile range (IQR), correlation results with Pearson correlation coefficient ($r$), and survey results with median ($Med$) and interquartile range ($IQR$). Performance improvements refers to any change above 0\%. }.

\begin{table}[!h]
\centering
\renewcommand{\arraystretch}{1.3}
\resizebox{\textwidth}{!}{
\begin{tabular}{p{1.1cm}p{1.6cm}p{1.6cm}p{1.2cm}p{1.5cm}p{1.5cm}p{1.7cm}p{1.6cm}p{1.0cm}p{1.6cm}}
\hphantom{h} \newline Metric & Correspond. (1-7) & Arm Travel  (meter) & Duration (ms) & Straightness  (0-1) & Peak Speed (m/s) & Time to Peak Speed  (ms) & Peak Speed to Target (\%) & Fitts ID & Throughput (bits/s) \\
\toprule
Distance & \cellcolor{g6}3.97 {\color{sd} 1.9} & \cellcolor{g6}0.28 {\color{sd} 0.1} & \cellcolor{g6}1349 {\color{sd} 370} & \cellcolor{g6}0.84 {\color{sd} 0.1} & \cellcolor{g6}0.83 {\color{sd} 0.4} & \cellcolor{g6}569 {\color{sd} 264} & \cellcolor{g6}59\% {\color{sd} 10} & \cellcolor{g6}2.74 {\color{sd} 0.6} & \cellcolor{g6}2.19 {\color{sd} 0.8} \\ 
Speed & \cellcolor{g6}4.20 {\color{sd} 1.9} & \cellcolor{g6}0.28 {\color{sd} 0.1} & \cellcolor{g6}1302 {\color{sd} 358} & \cellcolor{g6}0.83 {\color{sd} 0.1} & \cellcolor{g6}0.88 {\color{sd} 0.4} & \cellcolor{g6}532 {\color{sd} 234} & \cellcolor{g6}60\% {\color{sd} 10} & \cellcolor{g6}2.74 {\color{sd} 0.6} & \cellcolor{g6}2.27 {\color{sd} 0.8} \\ 
Time & \cellcolor{g6}4.62 {\color{sd} 1.7} & \cellcolor{g6}0.28 {\color{sd} 0.1} & \cellcolor{g6}1321 {\color{sd} 370} & \cellcolor{g6}0.83 {\color{sd} 0.1} & \cellcolor{g6}0.85 {\color{sd} 0.4} & \cellcolor{g6}543 {\color{sd} 242} & \cellcolor{g6}60\% {\color{sd} 9} & \cellcolor{g6}2.74 {\color{sd} 0.6} & \cellcolor{g6}2.24 {\color{sd} 0.8} \\
\toprule
\bottomrule
\end{tabular}}
\caption{Means (black numbers) and standard deviation (grey numbers) across each tested metric (time, speed, distance).}
\label{table:study-condition-means}
\end{table}

\subsection{RQ1: Self-regulation with Performance Metrics}
Following our instruction while receiving time-, peak speed, and distance-based dynamic feedback, participants changed their performance in terms of movement time and straightness. 
It varied between participants whether their performance improved or worsened from the dynamic feedback -- the majority of participants made straighter movements by 0-5\% (15/24, $M$=$1\%$, $SD$=$1.2\%$) and achieved higher peak speeds by 0-41\% (14/24, $M$=$12\%$, $SD$=$9.7\%$), while less than half of participants improved their completion time by 0-6\% (9/24, $M$=$3\%$, $SD$=$2\%$). The majority of players (21/24) improved their performance to one of the three metrics, except for three participants, whose performance with the metric worsened compared to their overall average. Performance differences shown in Fig.~\ref{fig:bar-performance-metric} tested with repeated measures ANOVA revealed significant changes between distance and peak speed conditions in terms of {{\color{rv1} movement time ($p$=$0.013$, $\eta^2$=$0.013$), and straightness ($p$=$0.040$, $\eta^2$=$0.010$), and peak speed ($p$=$0.021$, $\eta^2$=$0.014$). 
However, Bonferroni post hoc tests of peak speed did not reveal significantly different pairs}.
Players' overall average improvement fluctuated between -5\% and +11\% on average ($M$=$2\%$, $SD$=$4\%$).

\begin{figure}[!h]
\captionsetup{font=footnotesize}
\centering
\hspace{-5pt}
\caption*{Performance Measure Comparison across Metric Types}
\includegraphics[width=4.5cm]{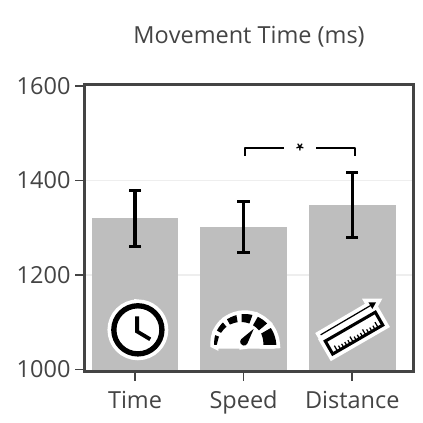}
\hspace{-5pt}
\includegraphics[width=4.5cm]{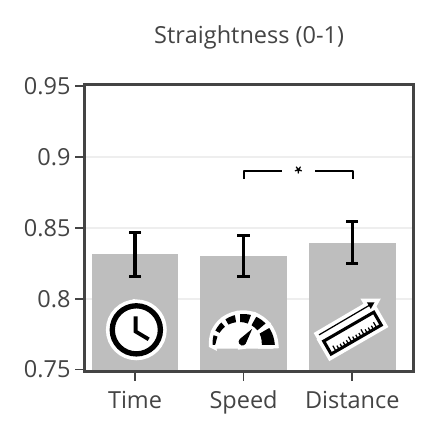}
\hspace{-5pt}
\vspace{3pt}
\includegraphics[width=4.5cm]{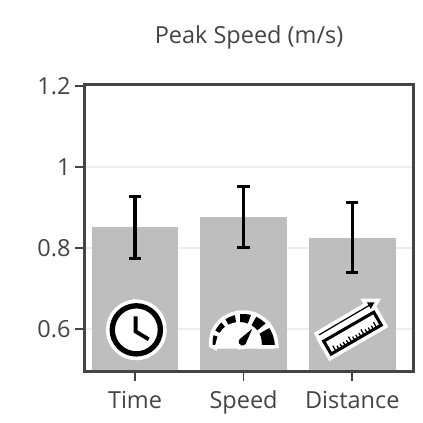}
\hspace{-5pt}
\caption{Performance measurement comparison for each metric type, with significant pairs in the Bonferroni post hoc test highlighted with an asterisk ($p<0.05$*) and error bars indicating 95\% confidence intervals.}
\label{fig:bar-performance-metric}
\Description{Performance Comparison across Metric Types. Three bar charts compare time, speed and distance metrics, across movement time, straightness, and peak speed. Black whiskers indicate the error of every bar, while black clamps annotated by stars indicate significant differences. Movement time and straightness show significant differences between speed and distance bars, with a higher value for distance. Peak Speed show speeds around 0.8 to 0.9 meters per second but no significance.}
\end{figure}

Linear regression across each counterbalanced session suggested that people learned over time to lower their mean completion time by 12\% ($\beta$=$-12.25$, intercept=$1335.98$), increase their mean peak speed by 4\% ($\beta$=$0.04$, intercept=$0.81$), but only marginally improved overall mean straightness by 0.02\% ($\beta$=$0.002$, intercept=$0.831$).
We normalized measures of completion time, straightness and peak speed and summed them up to form an overall performance measure.
Players' overall performance varied, approximating a normal distribution, where best performers performed 20\% better than the mean and 40\% better than the worst observed performance.
People's overall ability to adapt depended weakly on their overall performance, according to a spearman correlation test bordering significance (r = 0.23, p < 0.04797). 
We checked whether differences in participant performance followed differences in their ratings of correspondence. The Spearman correlation test showed a low correlation to action arm travel ($r$=$0.21$, $p$=$0.07$) and straightness ($r=$$-0.22$, $p$=$0.07$) and no correlation to other performance variables ($r$<$0.15$, $p$>$0.05$). Participants rated correspondence of time-based metrics {{\color{rv1} ($Med$=$5$, $IQR$=$3$)} significantly higher than distance-based metrics {{\color{rv1}($Med$=$4$, $IQR$=$3.25$)}, according to a Conover's F post hoc test. They did not rate correspondence with speed-based metrics {{\color{rv1}($Med$=$4$, $IQR$=$3$)} significantly different from time-based metrics ($p$=$0.063$) or distance-based metrics ($p$=$0.445$).
Correspondence positively correlated to most experiential measurements including how encouraging participants felt at moderate level ($r$=$0.51$-$0.72$, $p$<$0.001$), where the highest observed correlation was to ratings of feedback appeal ($r$=$0.72$, $p$<$0.001$). Correspondence correlated weakly to ratings of distraction ($r$=$0.26$, $p$=$0.03$). 

When questioned, no participant reported predicting the pseudo-randomized activation pattern with full confidence. We did not find significant effects of age, gender, glasses, VR sickness, or overall experience.
Supplementary Material 2 provides fully detailed test results from our ANOVA and post hoc tests. {\color{rv1} In terms of learning effects, visual trend line analysis within each session showed that participants decreased their average time to reach peak speed by 30ms ($\beta$=$-3.3ms$), and reduced their overall average movement times by 110ms ($\beta$=$-7.9ms$). Their throughput increased by 0.09 bits/s ($\beta$=$0.01bits$/$s$) -- less than 0.05\% -- and their straightness increased by 0.02 ($\beta$=$0.002$).}

\subsection{RQ2: Impressions of Cognitive Factors in Self-regulation}
Based on Likert scale ratings and the post-experiment interview, we analyzed participants’ experiences of self-regulation with dynamic feedback across the dimensions outlined in our conceptual model in Fig.~\ref{fig:conceptualspace} in terms of legibility, correspondence, preference, encouragement and distraction.  Violin plots in Fig.~\ref{fig:likert-performance-feedback} visually summarize how people rated the continuous trail, checkmark, and heatmap feedback, after experiencing them in that order.

\begin{figure}[!h]
\captionsetup{font=footnotesize}
\centering
\hspace{-5pt}
\caption*{Likert Scale Responses Per Feedback Types}
\includegraphics[width=3.67cm]{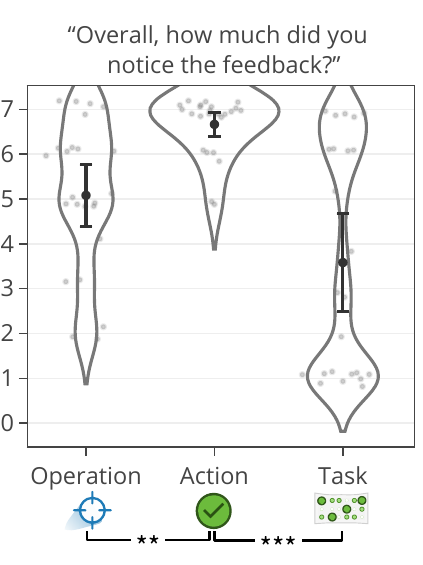}
\hspace{-5pt}
\includegraphics[width=3.67cm]{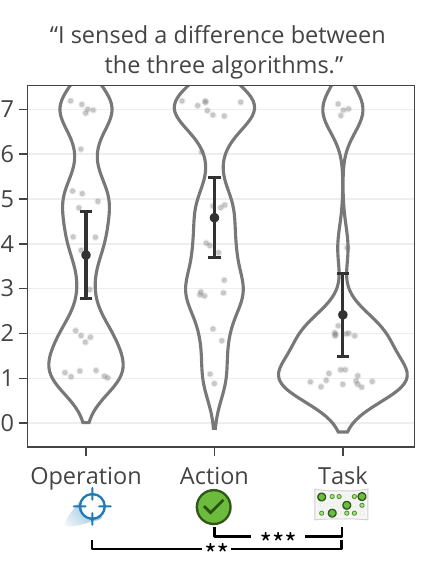}
\hspace{-5pt}
\includegraphics[width=3.67cm]{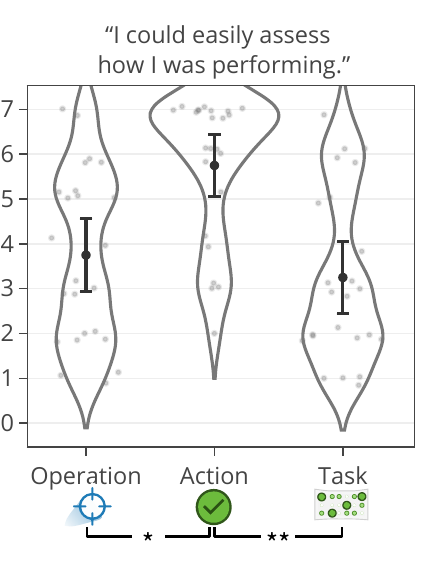}
\hspace{-5pt}
\includegraphics[width=3.67cm]{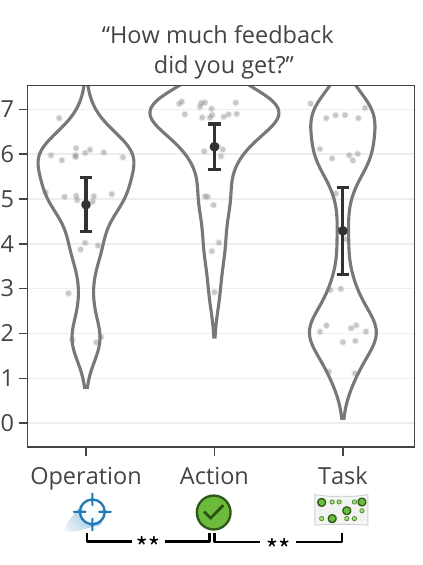}
\hspace{-5pt}
\includegraphics[width=3.67cm]{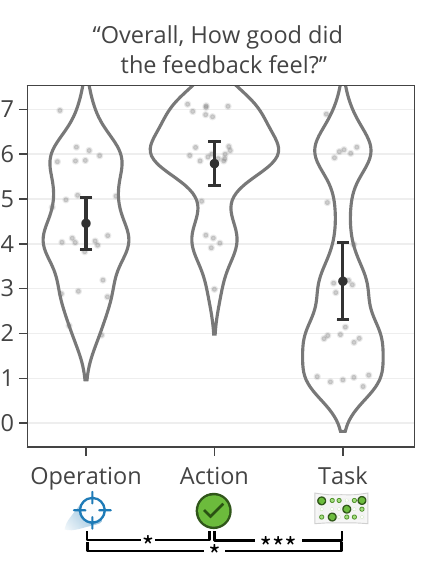}
\hspace{-5pt}
\vspace{5pt}
\includegraphics[width=3.47cm]{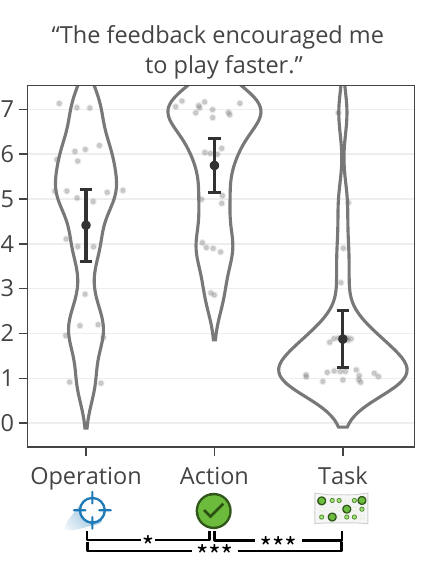}
\hspace{-5pt}
\includegraphics[width=3.47cm]{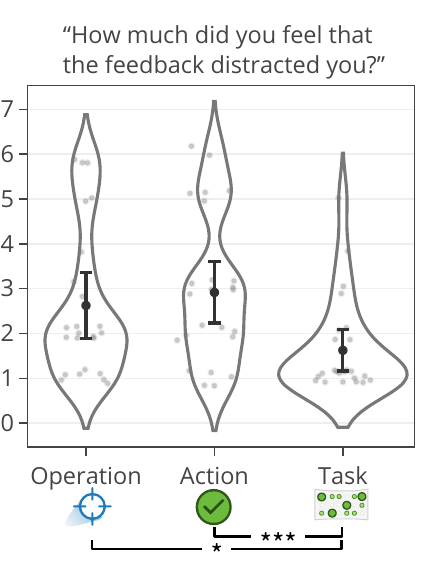}
\hspace{-10pt}
\caption{Measurement comparison for each feedback type, with significant pairs in the Friedman post hoc test highlighted with asterisks ($p<0.05$*, $p<0.01$**, $p<0.001$***). {\color{rv1} Measurements were derived from the self-regulation model in Fig.~\ref{fig:conceptualspace} (see Table~\ref{tab:experiential-survey}).}}\label{fig:likert-performance-feedback}
\Description{Likert Scale Responses Across Feedback Types. Seven violin plots titled Likert scale responses per feedback type. Grey dots depict participants' ratings, while black dots with error bars indicate the mean ratings.  Clamps and asterisks are shown under each compared feedback type, indicating significant differences. The first graph asks "Overall, how much did you notice the feedback?" Action feedback is highest rated and significantly different compared to task and operation feedback, task feedback shows a bimodal distribution. The graph "I sensed a difference between the three algorithms" depicts action feedback highest and significantly different with more spread on operation and task. The graph "I could easily assess how I was performing" shows action feedback with highest rating, significantly different from operation and heatmap feedback. The graph "How much feedback did you get" shows a bimodal distribution in the task feedback ratings, and the action feedback is rated highest. The graph "Overall, how good did the feedback feel?" shows significant differences between all ratings, where action feedback is rated highest and task feedback lowest. The graph "The feedback encouraged me to play faster." shows very low score to task feedback at less than 3, while actiona and operation feedback is higher than four, all significantly different from one another. The graph "How much did you feel that the feedback distracted you?" show significant differences between task feedback, compared to operation and action feedback, with values three or less across all feedback types.}
\end{figure}

\subsubsection{Perceptual Legibility}
People noticed checkmark feedback significantly more {{\color{rv1}($Med$=$7$, $IQR$=$0.25$, $\eta^2$=$0.33$)} than both trail {\color{rv1}($Med$=$5$, $IQR$=$1.25$}, $p$=$0.006$) and heatmap feedback {{\color{rv1}($Med$=$3$, $IQR$=$5$}, $p$<$0.001$), scoring the heatmap feedback binomially. 
The majority of participants paid much attention to the visual aspect of the feedback. They described that the blue trail following the cursor became ``longer'', ``thicker'', ``clearer'', and changed in opacity {\color{Highlight}(transparency)}.
They noticed that checkmarks became greener when they did better and had different opacity. 
To four participants, the heatmap felt overwhelming: \textit{``I got very overwhelmed by the feedback. Too much sound and visual at once.''} (P4).
Half of the participants (11/24) commented on the changes in sound pitch based on their performance, which was present in all three feedback sessions. Some participants described the blue trail's sound as a continuous ``whooshing'' or ``whirring'' sound: \textit{``As I move the cursor, it would grow a tail and make a whirring noise based on my timing/speed''} (P6).
Few participants (5/24) commented on the presence of haptic feedback, describing it as vibrations timed to the visual and auditory feedback.

\subsubsection{Contrasting Legibility}
People had a significantly harder time differentiating between performance metrics with heatmap feedback {{\color{rv1}($Med$=$1.5$, $IQR$=$1$, $\eta^2$=$0.35$)} than with trail {{\color{rv1}($Med$=$4$, $IQR$=$3.5$}, $p$=$0.007$) and checkmark feedback {{\color{rv1}($Med$=$4.5$, $IQR$=$4$}, $p$=$0.001$). Participant scores were spread, except for heatmap feedback where the vast majority scored it low (>3). All 24 participants remarked that they sensed being measured through the metrics, although they could not always tell how. 
Six participants stressed that when the metric did not feel right, the feedback could feel unclear or weird: \textit{``The first one [speed] seemed inverted [e.g. grey when fast and green was slow], the second one [time] felt very accurate and the third one [distance] felt like it didn't make much sense. The second one [time] felt really good and accurate but the two others were weird.''} (P11).
Three participants commented that, for them, their ability to differentiate it depended on the metric used to drive it.

\subsubsection{Analytical Legibility} 
People found it significantly easier to comprehend checkmark feedback {{\color{rv1}($Med$=$6.5$, $IQR$=$2.25$, $\eta^2$=$0.22$)}, than trail {{\color{rv1}($Med$=$3.5$, $IQR$=$3$}, $p$=$0.026$) or heatmap feedback {{\color{rv1}($Med$=$3$, $IQR$=$3$}, $p$=$0.002$). Four participants remarked during interview that the checkmarks were best to clarify metric differences.
Participants remarked the feedback was reacting to their timing, speed and to being fast.
Four participants commented that the trail-based feedback was not as intuitive to hint at their performance: \textit{``It became longer I think, I did not realize how long. Sometimes it was completely gone. [..] It wasn't clear how it changed.''} (P16).
The heatmap feedback was hard to comprehend and confusing to many participants (12/24): \textit{``I saw it after the task, which was kind of confusing the first time. It was very clearly showing where I was doing good or bad, but as it happens after the game, I don't know why I did bad or good.''} (P5).
Participants stated that comprehending heatmap feedback required them to recall how they hit the targets. This made it difficult for participants to determine whether they agreed how the metric reflected their performance. 

\subsubsection{Preferences}
Participants felt they received most feedback from checkmarks {{\color{rv1}($Med$=$7$, $IQR$=$1.25$, $\eta^2$=$0.20$)} compared to trail {{\color{rv1}($Med$=$5$, $IQR$=$2$}, $p$=$0.007$) and heatmap feedback {{\color{rv1}($Med$=$4.5$, $IQR$=$4.25$}, $p$=$0.009$). They did not agree whether heatmap feedback provided a lot or little feedback, scoring it binomially. Participants thought that the checkmark feedback felt best {{\color{rv1}($Med$=$6$, $IQR$=$1.25$, $\eta^2$=$0.37$))}, followed by trail feedback {{\color{rv1}($Med$=$4$, $IQR$=$2.25$}, $p$=$0.039$), and heatmap {{\color{rv1}($Med$=$2.5$, $IQR$=$3.5$}, $p$<$0.001$). The checkmarks were complimented by 16 participants as clear, easy to follow, accurate, and natural: \textit{``It made more sense to me, it's almost like pushing a button thanks to the short sound. The color change was a great visual feedback.''} (P8). 
The blue trail was described as satisfying, visually pleasing, and helpful: \textit{``I felt [the trail] helped explain how fast I was, especially the first one [speed], was very concrete.'' (P18)}
Two participants commented that the trail gave them a better overview and indication of where their cursor was. 
Three participants were not happy with the trail's sound: \textit{``The sound was confusing me but the visual worked very well, making it easier to see the cursor on the screen.''} (P8)
The majority preferred checkmark feedback (14/24), while others preferred trail feedback (6/24), and four participants indicated no preference. 
No participants preferred heatmap feedback, some commenting that it had resemblance to a ``scoreboard'' (P12), or being ``graded'' (P8) because they had no opportunity to action upon the feedback while playing.
Still, some participants shared positive remarks, saying the heatmap felt like a big reward (3/24), and that its timing gave them space to focus when playing (2/24): \textit{``It felt nice to sit and look at an overview of how you did, showing the areas you did well or not.''} (P2).

\subsubsection{Encouragement and Distraction}
Participants felt the checkmark feedback encouraged them most {{\color{rv1}($Med$=$6$, $IQR$=$2.25$, $\eta^2$=$0.57$)} compared to trail {{\color{rv1}($Med$=$5$, $IQR$=$3.25$}, $p$=$0.035$))}} or heatmap feedback {{\color{rv1}($Med$=$1$, $IQR$=$1$}, $p$<$0.001$)).Participants rated heatmap feedback as least distracting {{\color{rv1}($Med$=$1$, $IQR$=$1$, $\eta^2$=$0.24$))},compared to trail {{\color{rv1}($Med$=$2$, $IQR$=$2.25$}, $p$=$0.042$))}} and checkmark feedback {{\color{rv1}($Med$=$3$, $IQR$=$1.5$}, $p$=$0.001$))}}. One participant remarked: \textit{``I did not feel taken out of the game during [the heatmap] and didn't need to think about feedback while playing.''} (P2).  
Checkmarks were both the most encouraging and most distracting. Four participants stated that checkmarks disrupted their flow, made them feel anxious, or could throw them off when feedback was inaccurate.
The time it took participants to reach their peak speed with checkmarks was on average 580ms ($SD$=$112ms$), which was 6\% higher than the overall average time to reach peak speed ($M$=$548ms$, $SD$=$107ms$).

\section{Discussion}
The metrics' measurable impact on user performance, suggested that most users, provided with knowledge of the metric, could improve their performance through our non-textual dynamic feedback.
Yet our results also highlight a potential performance penalty introduced by dynamic feedback in some cases. For example, the checkmark feedback shown at the end of each action was simultaneously the most preferred and most distracting, creating a measurable performance cost - an additional 6\% time to  reach peak speed.
Motor learning studies have often suggested that learners, once having learned to perform, benefit from the withdrawal of the feedback or less frequent feedback~\citep{sigrist_augmented_2013} -- highlighting the value of understanding potential performance costs of feedback.
However, unlike motor learning, games face extra considerations when designing augmented feedback -- motor learning primarily removes feedback to optimize performance and avoid over-reliance~\citep{sigrist_augmented_2013}, while game designers use it to create variability, rewards ,and balance game challenge~\citep{Schell_2019}.

Across feedback with different timing we identified positive and negative impacts on player performance and their sense of correspondence.
The end-of-action checkmark feedback was the overall most encouraging and preferred. 
However, checkmark feedback was simultaneously the most demanding to decode in this experimental setting that dual-tasked participants with play their best while evaluating the feedback. Checkmark feedback yielded the lowest peak speed and phase of movement from peak speed to target. 

Our study of functional dynamic feedback adds to the understanding of the design tension between encouraging (e.g juicy) feedback that users appreciate, and the cost from possible  distractions, similar to the performance penalties found in \citeauthor{Hicks_2019}'s evaluation of juicy non-functional feedback ~\citep{Hicks_2019}. 
Our study showed a potential bias in participants' internal sense of performance -- users favoring checkmark feedback despite its measurable cognitive costs may indicate that they favored positive rewards over real performance improvements. 
But our study was not designed to answer to what degree these improvements were due to legibility difficulties in decoding and evaluating the feedback or the differences in how the metrics fostered better performance.
The bimodal distributions of the quantity and perceivability of the heatmap feedback ratings in Fig.~\ref{fig:likert-performance-feedback} suggested that the users might have interpreted or emphasized the questions differently. Participants, who rated the quantity and perceivability of heatmap low, argued that they did so since no feedback was present during the task.

Our conceptual model (Fig.~\ref{fig:conceptualspace}) theorized that feedback influences performance based on the extent to which individuals feel a sense of correspondence, encouragement, or distraction. In this study, distraction from feedback determined users' performance more than its correspondence or encouragement. While participants preferred feedback with higher correspondence, its correlation to performance was negligible.

\subsection{Contextualization of the work}
\citeauthor{wensveen_interaction_2004}'s notion of naturally coupled feedback connects to notions like natural mapping and intuitive interaction, also discussed in the context of designing control interfaces, like motion-controlled gaming~\citep{McEwan_2020,Skalski_2011}. They also utilize a notion of correspondence, which in contrast to this study,  refers to how naturally the physical control interface maps to virtual effects in the game world at different levels~\citep{Skalski_2011}, and quantifies perceived naturalness~\citep{McEwan_2020}.

\citeauthor{Kao_2024}'s research on low-level action-feedback game loops found that excessive amplification can weaken the causal link between action and response, reducing legibility and user understanding~\citep{Kao_2024}.
They tied the efficacy of feedback to curiosity as an underappreciated engagement driver, of continued gaming~\citep{Kao_2024} -- the way players can explore the different possible states of animations, provided a novelty effect. \citeauthor{Kao_2024} highlighted in particular the importance of providing players agency over the exploration, for example by tying feedback to skill rather than random states~\citep{Kao_2024}.  This makes feedback dynamics an important agency factor, which may be positively affected when dynamic feedback is tied to measures of user performance. 
However, effectance is typically framed from the perspective that games must respond to players' input immediately to evoke strong and continuous experiences of effectance~\citep{Klimmt_2007} -- studies need to investigate how effectance is affected by loose or tight temporal couplings.
Our study's framing differs from \citeauthor{Kao_2024}'s study, because all possible feedback states (how they might vary in size, opacity, etc.) were explained to users upfront, and because we asked users to make active assessments of what they saw -- not only judging causality but judging the extent to which they felt the system's feedback matched their own. 

Dynamic feedback is related to corrective outcome feedback in instructional learning~\citep{Johnson_2017}, which serves to direct users' attention to key task aspects, making learners aware of the gap in performance, and apply correction strategies to progress~\citep{Shute_2008}. 
Research within learning games use other models to represent cognitive processing of feedback, including notions like \textit{external} and \textit{internal} attention, to describe how incoming external feedback is organized and integrated with internally generated information~\citep{Gauthier_2022}. \citeauthor{Gauthier_2022} described how feedback's \textit{salience} is important for users to attend to feedback - which is controlled by its content, timing and modality\citep{Gauthier_2022}. Compared to studies of learning games where feedback operates at larger time scale, this study dealt with feedback and its ability to induce self-regulation in low-level action loops, functioning on a micro-level from a self-regulatory learning perspective. {\color{rv1} Likewise, fast-paced game genres, such as action or racing games, the need for immediate and clear feedback becomes crucial, as players rely on real-time information to make split-second decisions. In contrast, slower, strategy-based games may benefit from more gradual feedback mechanisms, allowing players to reflect on their actions over longer durations.}

{\color{rv1} 
For cue research, the design space (Section~\ref{sec:dimensionsfeedback}) contrasts existing cue frameworks~\citep{Dillman_2018}, which focus more on the feedback's role (e.g. what it prompts users to do), rather than the dimensions of design.
}

\subsection{Limitations}
This study utilized three scenarios trail feedback, checkmarks and heatmaps, to quantify how people adapted performance based on specific metric with full knowledge of its function. The study was designed to compare metrics and not designed to quantify performance improvements from specific feedback types. {\color{rv1}The absence of a baseline condition limited what was known about players' performance without feedback. By design, the study therefore cannot provide absolute benefits or costs of dynamic feedback. Designing dynamic feedback always involves multiple design choices in terms of expression, location, modality and timing, which can impact participants' performance, encouragement, and distraction. Quantitative results and participant's responses will reflect the unique interplay of design choices in the study's feedback and the individualized feedback adaption system.  The study did not not utilize pre-study trials, which limits the accuracy of the initial feedback.}
} 
We have hinted at potential player preferences in our qualitative research, but leave it for future work to quantitatively compare the added benefit of trail, checkmark and heatmap feedback.

This study has sought to exemplify a systematic exploration of dynamic feedback in a pointing game context. Our results are based on a target-to-target pointing task commonly found in shooting games but might not apply
in games with longer self-regulation cycles leaving  more reflection time during turn-taking or disruptions~\citep{Miller_2024}. 

Lower ratings of correspondence for the straightness metric may stem from the use of the word 'fast,' whose connection to making accurate trajectories may have been unclear to participants. During pilot testing, 'fast' was clearer, while terms like 'better performance' lacked precision. Despite this, participants still acquired targets more accurately when given feedback based on the straightness metric.

\subsection{Future work}
Our study opens new avenues for exploring dynamics of feedback and its temporal placement in user interactions. Other metrics could be evaluated to understand participants' ability to adapt to them, such as metrics that reward fast acquisition of targets (i.e., the corrective movement phase in \citeauthor{meyer_speedaccuracy_1990}'s model) or rewards reaction time to the presence of the next target (visual search~\citep{meyer_speedaccuracy_1990}). 
Our goal was to design metrics that were easy to explain and test whether they could moderate play behavior. 
Future work could investigate how advanced metrics based on advanced models of human performance or artificial intelligence fare in encouraging specific play behaviors. 

Complex feedback that integrates multiple sources of information, risks being difficult to understand and predict, due to their \textit{black-box} nature ~\citep{adadi_peeking_2018}. This legibility issue has led to research in how systems can create explainable user interfaces by providing post hoc explanations, like exploratory data visualization to audit results~\citep{shneiderman_bridging_2020}. Our findings suggests that complex feedback systems will need additional design features (e.g. meaningful supplemental textual feedback) to reach enough legibility (or \textit{algorithmic transparency}~\citep{Schor_2024}).
Studies building upon our results could compare no-feedback scenarios, non-dynamic feedback, or measures participants twice per condition, to e.g. make it possible to measure players' adaption to task feedback. 
Similarly, explorations of how other feedback modalities might fare e.g., textual feedback overlays such as percentage-wise feedback (e.g. ``80\%'' as a distance from their best performance, which would be ``100\%'') or measurements  in standardized units (``8m/s'', ``2.3m'', ``4.1s''), or situational feedback such as effects of combos, streaks,  multipliers, or accumulative mechanics. 

{\color{rv1}
Our study results are based upon a simple pointing game in a VR environment, which offered experimental control and generalization. Our methodology could be applied for pointing in more complex game environments resembling off-the-shelf games like Fruit Ninja or Counter-Strike, which often feature ultimate goals (e.g. winning/loosing) and imperative goals (e.g. acquiring points/kills) within highly developed fictive worlds.
}

Our work and the conceptual model of interaction can be applied in future game scholarship, in, for example, motion-controlled VR games for entertainment and training applications, and beyond VR in HCI systems for which improvements in motor performance are a goal, such as improving eye-hand coordination~\citep{batmaz_improving_2022}, rehabilitation~\citep{Castillo_2024}, or exercise~\citep{Ioannou_2019}.
Exploring optimal dynamic feedback strategies may also be relevant for competitive electronic sport. Elite esport performers and coaches constantly look for means of achieving superior accuracy and time trial performance. Behavioral monitoring behavior and feedback plays an important role as part of cyclical self-regulation to optimize esport performance~\citep{Trotter_Sharpe_2023}. 
By interpreting player movements and providing interpreted feedback, games can take a supplemental role to coaching, regulating players' behavior to achieve a specific result~\citep{Cote_2006}.

Considering the limitations of the results and the design of this reported study, the authors are continuing to redesign and re-evaluate the metrics and feedback using a more elaborate, counterbalanced setup. This will allow us to more clearly disentangle the effects of feedback, instruction, and different learning and fatigue effects.

\subsection{Design Guidelines}
{\color{rv1} Based upon our results, we suggest the following design guidelines for feedback in pointing tasks.  When designing performance metrics, designers should focus on designing metrics for the tightest temporal coupling first, as they translate better to looser non-continuous feedback through aggregation than in reverse. 
\textbf{For minimal distraction and speed penalty}, metrics should utilize tight temporal coupling like operation feedback.
\textbf{For optimal legibility}, performance metrics should be tied to action feedback instead, but practitioners should be aware that it can create performance and distraction costs for subsequent actions, which may not be easily recognized by participants. Action feedback needs to be easy to decode and minimize interruptions.
When designing \textit{task-level feedback}, practitioners should be mindful of potentially low correspondence. To support players to improve, it should be coupled with action- or operation-level feedback.} 

{\color{rv1} Designers can choose to incorporate dynamic feedback as part of a cue, if the source metric information is meaningful for players' task. Wensveen's feedback dimensions may also help studies of cues to explore and explain possible mappings with greater detail. For example,~\citeauthor{Karaosmanoglu_2024}'s study of body posture cues~\citep{Karaosmanoglu_2024}, could situate their choice of cue design within a much larger design space than merely color. Likewise, experiential measures like correspondence could help them decide upon choice of metric to provide continuous body pose cues.
}  

\section{Conclusion}
Our study evaluated how players’ adapted their performance to specific performance metrics, visualized as dynamic feedback in a simple VR pointing task. 
{{\color{rv1}The study was scoped to quantitatively compare metrics, leaving form-to-form performance contrasts beyond scope.}
On average, participants improved their performance according to performance metrics of completion time, peak speed or movement straightness, which drove dynamic feedback displayed continuously, end-of-action and end-of-task. 
To achieve desired player behavior, game designers should carefully align performance metrics with a game's objectives, to support  players' reaching their potential -- e.g., use a peak speed metric if achieving higher peak speed is the goal.
However, users may prefer mappings that do not optimize performance.
Designers should also be aware of the performance cost feedback may induce -- although feedback shown at the end of each action may be preferred by players, {{\color{rv1} they also consider it distracting (e.g. impose a cognitive cost for subsequent actions.)}
Our results demonstrate 
how providing a conceptual self-regulation model can be used as a design tool to explain the effects of feedback collected in experiments and through playtesting. 
our work opens up a new avenue for exploring how basic dynamic feedback can be systematically studied and how it can impact player experience.

\balance{}

\bibliographystyle{ACM-Reference-Format}
\bibliography{references}

\end{document}